\documentclass[12pt,a4paper]{article}
\usepackage{amsmath,amssymb,mathrsfs,framed,esint,slashed}
\usepackage[colorlinks]{hyperref}
\usepackage{color,comment}
\usepackage[all]{xy}
\usepackage[color=yellow]{todonotes}

\usepackage{braket}
\usepackage{bbm}

\newtheorem{theorem}{Theorem}[section]

\newtheorem{remark}[theorem]{Remark}

\newcommand{\rem}[1]{}

\newcommand{\de}{{\rm d}}

\newcommand{\bq}{{\boldsymbol{q}}}
\newcommand{\bv}{{\boldsymbol{v}}}
\newcommand{\bp}{{\boldsymbol{p}}}

\newcommand{\bm}{{\boldsymbol{m}}}

\newcommand{\bx}{{\boldsymbol{x}}}
\newcommand{\bsA}{{\boldsymbol{A}}}

\newcommand{\bsr}{{\boldsymbol{r}}}
\newcommand{\bsx}{{\boldsymbol{x}}}

\newcommand{\bsmu}{{\boldsymbol{\mu}}}
\newcommand{\bsmubar}{{\boldsymbol{\bar{\mu}}}}
\newcommand{\bomega}{\boldsymbol{\omega}}

\newcommand{\bmu}{\boldsymbol{\mu}}
\newcommand{\bbeta}{\boldsymbol{\eta}}
\newcommand{\bn}{{\boldsymbol{n}}}

\newcommand{\bw}{{\boldsymbol{w}}}
\newcommand{\br}{{\boldsymbol{r}}}
\newcommand{\bs}{{\boldsymbol{s}}}

\newcommand{\by}{{\boldsymbol{y}}}
\newcommand{\bA}{{\boldsymbol{A}}}

\newcommand{\bB}{{\boldsymbol{B}}}
\newcommand{\bJ}{{\boldsymbol{J}}}

\newcommand{\bu}{{\boldsymbol{u}}}

\newcommand{\bbR}{\mathbb{R}}
\newcommand{\Dbar}{\bar{D}}

\newcommand{\beq}{\begin{equation}}
\newcommand{\eeq}{\end{equation}}
\newcommand{\bal}{\begin{align}}
\newcommand{\eal}{\end{align}}
\newcommand{\ben}{\begin{eqnarray}}
\newcommand{\een}{\end{eqnarray}}

\renewcommand{\contentsname}{}

\numberwithin{equation}{section}
\numberwithin{figure}{section}

\begin{document}

\title{Geometry of nonadiabatic quantum hydrodynamics
}
\author{Michael S. Foskett$^{1,2}$, Darryl D. Holm$^3$, Cesare Tronci$^{1,2}$
\smallskip
\\ 
\small
$^1$\it Department of Mathematics University of Surrey, UK
\\
\small
$^2$\it Mathematical Sciences Research Institute, Berkeley CA, USA
\\
\small
$^3$\it Department of Mathematics, Imperial College London, UK}
\date{\tiny \vspace{-.85cm}}
\maketitle

\begin{abstract} 
\small
The Hamiltonian action of a Lie group on a symplectic manifold induces a momentum map generalizing Noether's conserved quantity occurring in the case of a symmetry group. Then, when a Hamiltonian function can be written in terms of this momentum map, the Hamiltonian is called  `collective'.
Here, we derive collective Hamiltonians for a series of models in quantum molecular dynamics for which the Lie group is the composition of smooth invertible maps and unitary transformations. In this  process, different  fluid descriptions emerge from different factorization schemes for either the wavefunction or the density operator.
After deriving this series of quantum fluid models, we regularize  their Hamiltonians for finite $\hbar$ by introducing local spatial smoothing. In the case of standard quantum hydrodynamics, the $\hbar\ne0$ dynamics of the Lagrangian path can be derived as a finite-dimensional canonical Hamiltonian system for the evolution  of singular solutions called `Bohmions', which follow Bohmian trajectories in configuration space. For  molecular dynamics models, application of the smoothing process to a new factorization of the density operator leads  to a finite-dimensional Hamiltonian system for the interaction of multiple (nuclear) Bohmions and a sequence of electronic quantum states.

\end{abstract}

\vspace{-1.05cm}

{\scriptsize
\contentsname
\tableofcontents
}

\addtocontents{toc}{\protect\setcounter{tocdepth}{2}}

\section{Introduction}\label{Intro-sec}

\subsection{Factorized wave functions in quantum molecular dynamics}

Quantum molecular dynamics deals with the problem of solving the {\it molecular Schr\"odinger equation}
\beq
i\hbar\partial_t \Psi=\big(\widehat{T}_n+\widehat{T}_e+\widehat{V}_n+\widehat{V}_e+\widehat{V}_I\big)\Psi=:\widehat{H}\Psi
\,,
\label{mol-eq}
\eeq
which governs the quantum evolution for a set of nuclei interacting with a set of electrons. In the equation above, $\Psi(\{\br_k\},\{\boldsymbol{x}_{l}\},t)$ is called the {\it molecular {wave function}}. The notation is such that $\{\br_k\} =  \{\br_k : k=1,\cdots,N_n\}$ and $\{\boldsymbol{x}_{l}\}=\{\bx_{l} : l=1,\cdots,N_e\}$ denote, respectively, $N_n$ nuclear and $N_e$ electronic coordinates. Each $\br_k$ corresponds to a nucleus of mass $M_k$ whilst all electrons have the same mass $m$. The notation $\widehat{T}$ and $\widehat{V}$ in \eqref{mol-eq} refers to the kinetic energy and potential energy operators, while the subscripts $n$ and $e$ denote nuclear and electronic energies, respectively. The subscript $I$ refers to the interaction potential between nuclei and electrons.  More explicitly, one has
\beq
\widehat{H}=-\frac{\hbar^2}{2}\sum_{k=1}^{N_n}\frac1{M_{k}}\Delta_{\br_k}-\frac{\hbar^2}{2m}\sum_{l=1}^{N_e}\Delta_{\bx_{l}}+{V}_n(\{\br_k\})+ {V}_e(\{\boldsymbol{x}_{l}\})+ {V}_I(\{\br_k\},\{\boldsymbol{x}_{l}\})
\,,
\label{mol-eqn}
\eeq
where the potentials are of Coulomb type \cite{Marx}. Without loss of generality, in this paper we shall consider only two particles, one nucleus and one electron, with coordinates denoted by $\br$ and $\bx$, respectively.

As the molecular Schr\"odinger equation is practically intractable for standard computational methods, a series of different closures and approximations for extracting its dynamics have been developed over almost a century. Since the work of Born and Oppenheimer \cite{born1927quantentheorie} in 1927, many efforts have been devoted to going beyond the adiabatic approximation in molecular dynamics, e.g., in the Jahn-Teller transition \cite{WoCe2004}.
In the standard approach, one separates out the nuclear kinetic energy term by writing the molecular Hamiltonian operator as the sum, 
\beq
\widehat{H}=\widehat{T}_n+\widehat{H}_e
\,.\label{mol-Ham-sum}
\eeq
Here, $\widehat{H}_e$ is called {\it electronic Hamiltonian}. Since $\widehat{H}_e$ also includes the potential energy operators  $\widehat{V}_n$ and $\widehat{V}_I$ (both depending on $\br$), one may write $\widehat{H}_e=\widehat{H}_e(\br)$. The Hamiltonian operator $\widehat{H}_e(\br)$ acts on the electronic Hilbert space $\mathscr{H}_e$, identified with the space of $L^2-$functions of the electronic coordinates, $\boldsymbol{x}$, which depend parametrically on the nuclear coordinates, $\br$. Thus, its eigenvalue equation reads
\[
\widehat{H}_e(\br)\phi_n(\boldsymbol{x};\br)=E_n(\br)\phi_n(\boldsymbol{x};\br)
\,.
\]
At every point $\br$, the eigenvectors $\phi_n(\boldsymbol{x};\br)$ provide an orthonormal frame in $\mathscr{H}_e$ and the level sets of the eigenvalues $E_n(\br)$ comprise hypersurfaces in the nuclear coordinate space, called {\it potential energy surfaces} (PES) \cite{Marx,Wyatt}. As customary in the chemical physics literature, for simplicity, here we assume a discrete electronic spectrum.
Formally, one can solve the molecular Schr\"odinger equation by writing the so-called {\it Born-Huang expansion} \cite{born1955dynamical}
\[
\Psi(\br,\boldsymbol{x},t)=\sum_{n=0}^\infty \Omega_n(\br,t)\phi_n(\boldsymbol{x};\br)
\,.\]
At this point, one can proceed by writing the equations for the coefficients $\Omega_n(\br,t)$, which in chemical physics are interpreted as {\em nuclear {wave function}s}. 

Historically, various additional approximations have been made that have treated the nuclei as classical particles. In particular, the lowest-order truncation of the Born-Huang expansion is the Born-Oppenheimer (BO) approximation, \cite{born1927quantentheorie}
\[
\Psi(\br,\boldsymbol{x},t)=\Omega_0(\br,t)\phi_0(\boldsymbol{x};\br)
\,.
\]
In the BO approximation, the electrons remain in the ground state identified with the lowest nuclear eigenvalue $E_0(\br)$ (\emph{adiabatic hypothesis}). Within the physical chemistry community, it is widely accepted that stable molecular configurations correspond to the minima of the lowest energy PES, $E_0(\br)$, see, e.g., \cite{Marx}. 

Despite the several successes of the BO approximation, the adiabatic hypothesis appears to be too restrictive for realistic computer simulations. Consequently, a great deal of work has been devoted to formulating mathematical models for nonadiabatic {dynamics \cite{Baer,deCarvalho}. Some of these approaches still exploit the Born-Huang expansion, while others introduce a fully time-dependent ansatz for the molecular {wave function} $\Psi$. Among the most acknowledged models for capturing nonadiabatic effects, the {\em mean-field model} is probably the simplest. The standard mean-field {factorization} ansatz is given by
\beq
\Psi(\br,\boldsymbol{x},t)=\chi(\br,t)\psi(\boldsymbol{x},t)
\,,
\label{MF-intro}
\eeq
where both $\chi$ and $\psi$ are {wave function}s (in their corresponding Hilbert spaces; respectively, $\mathscr{H}_n$ and  $\mathscr{H}_e$). After finding the wave equations for $\chi$ and $\psi$, semiclassical methods are typically applied to the nuclear {wave function}, $\chi$, thereby treating the nuclei as classical particles. Two different approaches are commonly used in dealing with {factorized} wave functions for molecular dynamics: 1) Frozen Gaussian wavepackets, \cite{Heller,Littlejohn2}, whose underlying geometric structure is based on coherent states \cite{Perelomov}; and 2) Bohm's hydrodynamic approach \cite{Bohm} which reduces the nuclear dynamics to a Hamilton-Jacobi equation. While the {geometry of the} first approach has been illustrated in \cite{OhLe,BLTr2015,OhTr2017} (see also \cite{deGosson} for related discussions), the present paper deals only with Bohm's hydrodynamic approach \cite{Bohm}.

While capturing nonadiabatic effects, the mean-field model does not adequately reproduce particle correlations between nuclei and electrons, \cite{Marx}. Thus, the mean-field model is apparently too simple to apply accurately to realistic situations in which correlations are important. Therefore, alternative methods have been developing during the past few decades. For example, in Tully's surface hopping algorithm \cite{Tully1,Tully2}, probability amplitudes are used to design a stochastic algorithm that enforces ``hopping'' between different energy levels $E_n(\br)$. In recent years, an augmented factorization scheme has also been proposed \cite{abedi2010exact,abedi2012correlated}, following earlier works by Hunter \cite{Hunter} and going back to von Neumann's book \cite{vonNeumann}. This approach is currently known as {\it exact {factorization}} (EF), which reads as follows
\beq
\Psi(\br,\boldsymbol{x},t)=\chi(\br,t)\psi(\boldsymbol{x},t;\br)\,.
\label{EF-intro}
\eeq
In the exact {factorization} approach, the electronic {wave function} $\psi$ depends parametrically on the nuclear coordinates $\br$; so, it can be regarded as a mapping from the nuclear coordinate space into the electronic Hilbert space $\mathscr{H}_e$. In this sense, the exact {factorization} provides a time dependent {generalization} of the BO approximation. Although the classical limit of the nuclear {wave function} $\chi(\br,t)$ could also have been taken by exploiting Gaussian wavepackets, the present work will investigate exact {factorization} by employing Bohm's hydrodynamic approach.  For an excellent review of the Gaussian wavepacket approach to nonadiabatic electronic effects, see \cite{Jo-DoIzmaylov2018}.

{\color{black}
This paper aims to investigate and compare the hydrodynamic approaches for both the mean-field models and the exact {factorization} ansatz in the context of geometric mechanics. Within this framework, separating out the nuclear kinetic energy  in the molecular Hamiltonian  \eqref{mol-Ham-sum} corresponds in the hydrodynamic approach to transforming into a Lagrangian coordinate  frame moving with the nuclei. 
Then, the momentum map associated to the evolution of the nuclear  wave function {\em collectivizes} in the sense of Guillemin and Sternberg, \cite{Guillemin,GuSt2013,LiMarle2012,marle2014,Marmo1985}. 
This means that equivariant momentum maps transform canonical Hamiltonian dynamics into motion on coadjoint orbits generated by the action of a Lie  group on the dual of its Lie algebra. 
Eventually,  Lie-Poisson reduction leads to a new hydrodynamic formulation of nonadiabatic dynamics in which {\em hyperbolicity is retained}, rather than setting $\hbar\to0$ and passing to the Hamilton-Jacobi equation. 
}

\begin{enumerate}

\item
The remainder of Section \ref{Intro-sec} introduces background material which links standard elements of quantum mechanics with 
familiar objects in the setting of geometric mechanics. The fundamental variational principles and symplectic Hamiltonian structure in nonrelativistic quantum mechanics appear in Section \ref{1.2-sec}. 

\item
The transformation to quantum hydrodynamics  is discussed in Section \ref{1.3}. In Section \ref{Bohmian-paths}, Bohmian trajectories \cite{Bohm} are reinterpreted as Lagrangian paths associated with the quantum hydrodynamic flow. Section \ref{QHE-Bohmions} {regularizes} the $\hbar\to0$ limit of standard quantum hydrodynamics by suitably applying a spatial smoothing operator to the fluid variables of both the collectivized Hamiltonian and the corresponding Lagrangian before taking $\hbar\to0$.
The resulting {smoothed} quantum fluid equations {are found to} admit singular solutions supported on delta functions. We call these singular solutions `Bohmions', because the delta functions on which they are supported move along Lagrangian paths of the {regularized} quantum fluid Hamiltonians.  Section \ref{densities} shows how the cold-fluid closure of Wigner distributions corresponds to a classical closure of mixed state dynamics arising from the Liouville-von Neumann equation.

\item
In the mean-field approximation of coupled nuclear and electronic systems, the wave function is separated into a product of two independent factors, as in equation \eqref{MF-intro}, above. Thus, the mean-field {factorization} of the wave function neglects the classical-quantum correlations between nuclei and electrons.  Section \ref{meanfield-sec} reviews the mean-field model and derives its quantum fluid representation in the geometric mechanics setting. 

\item
The {exact factorization} (EF) model \cite{abedi2010exact,abedi2012correlated,alonso2013comment,abedi2013response} captures some of the nuclear and electronic correlation effects which are neglected in the mean field approximation, by letting the electron wave function depend on the nuclear spatial parameters, as in equation \eqref{EF-intro}, above. Section \ref{EF-sec} discusses the EF model in both the wave function and density matrix representations, then derives its quantum fluid representation in the geometric mechanics setting.

\item In Section \ref{Bohmions} a new model is introduced by invoking a factorization ansatz at the level of the molecular density operator. Then, combining the classical closure of  nuclear mixed states with the smoothing process presented in Section \ref{QHE-Bohmions} leads to an entirely finte-dimensional Hamiltonian system for the interaction of nuclear Bohmion solutions with an ensemble of quantum electronic states. Two different finite-dimensional schemes are presented depending on whether the smoothing process is applied in the Hamiltonian or in the variational formalism.

\end{enumerate}

\subsection{Quantum Lagrangians, Hamiltonians, and momentum maps}\label{1.2-sec}

This section introduces  the standard setting of the Hamiltonian approach to quantum dynamics by focusing on the evolution of {\em pure quantum states}. Later sections of this work will introduce von Neumann's density operators and their evolution for mixed states. However, the present section considers only Schr\"odinger-type equations.

Among the most commonly used tools in chemical physics, the Dirac-Frenkel (DF) variational principle \cite{diracfrenkel} for the evolution of the {wave function} is expressed in terms of the symplectic  Hamiltonian structure of Schr\"odinger's equation. For a time-dependent quantum state $\psi(t)$ in the Hilbert space $\mathscr{H}$, the DF variational principle is expressed as a \textit{phase space Lagrangian}, which reads
\beq
\delta\int_{t_1}^{t_2}\langle\psi,i\hbar\dot\psi-\widehat{H}\psi\rangle\,\de t=0
\,.
\label{DFVarPrinc}
\eeq
Here, the bracket operation, {$\langle\,\cdot\,,\,\cdot\,\rangle$,  defines the real-valued pairing}
\beq
\langle\psi_1,\psi_2\rangle:=\operatorname{Re}\,\langle\psi_1|\psi_2\rangle
\,,
\label{braket}
\eeq
which is induced by the natural inner product $\langle\psi_1|\psi_2\rangle$  on $\mathscr{H}$, 
{given by
$
\langle\psi_1|\psi_2\rangle
=
\psi_1^\dagger\psi_2
=
\int \!\psi_1^*(\boldsymbol{x})\psi_2(\boldsymbol{x})\,\de^3x
$,
in which, e.g., $\psi_1^*$ denotes the complex conjugate of $\psi_1$, and $\psi_1^\dagger$ carries an implied integration.}

The Schr\"odinger equation $i\hbar\dot\psi=\widehat{H}\psi$ follows as the Euler-Lagrange equation for the DF variational principle in  \eqref{DFVarPrinc}, in which $\widehat{H}$ is the self-adjoint Hamiltonian operator constructed from the canonical operators $\widehat{Q}$ and $\widehat{P}$ (the so called {\it canonical observables}). Thus, $\widehat{H}=\widehat{H}(\widehat{Q},\widehat{P})$ and {$[\widehat{Q},\widehat{P}]=i\hbar\mathbbm{1}$}. Notice that, since $\widehat{H}$ is self-adjoint, the DF Lagrangian in \eqref{DFVarPrinc} is $U(1)-$invariant so that the condition $\|\psi\|_{L^2}^2=1$ is naturally preserved. This amounts to conservation of the  total probability. 
As presented in \cite{BLTr2015}, the Euler-Poincar\'e formulation \cite{HoMaRa1998} of pure state dynamics is derived from the DF variational principle upon letting $\psi(t)=U(t)\psi_0$ with $U(t)\in\mathcal{U}(\mathscr{H})$. Here, $\mathcal{U}(\mathscr{H})$ denotes the group of unitary operators on $\mathscr{H}$. In earlier years, this strategy was also exploited in \cite{KrSa,ShiRa} upon restricting $U(t)$ to be the unitary representation of a finite-dimensional Lie group. For example, if $U(t)$ is a representation of the Heisenberg group, substituting the ansatz $\psi(t)=U(t)\psi_0$ into the DF variational principle yields canonical Hamiltonian motion on phase space \cite{Shi}.

Notice that in \eqref{DFVarPrinc}, the functional $h(\psi)=\langle\psi,\widehat{H}\psi\rangle$ identifies the total energy of the system and, thus, it is deemed the Hamiltonian functional. The functional $h(\psi)$ is sometimes called {\it Dirac Hamiltonian}, to distinguish it from the Hamiltonian operator, $\widehat{H}$. {\color{black}Depending on the context, the operator $\widehat{H}$ and the functional $h(\psi)$ may both be called the `Hamiltonian'.} 
More general systems (such as the nonlinear Schr\"odinger equation) can be obtained by replacing $\langle\psi,\widehat{H}\psi\rangle$ by a suitable functional $h(\psi)$. In this case, the {normalization} condition $\|\psi\|_{L^2}^2=1$ must be incorporated as a constraint, \cite{Ohta}, as
\beq
\label{VarPrinc2}
\delta\int_{t_1}^{t_2}\!\big(\langle\psi,i\hbar\dot\psi\rangle-h(\psi)+\lambda(\|\psi\|^2-1)\big)\,\de t=0
\,,
\eeq
where $\lambda(t)$ is a real-valued Lagrange multiplier. For such constrained systems, the Euler-Lagrange equations yield the {\it projective Schr\"odinger equation} \cite{Kibble}
\beq
(\mathbbm{1}-\psi\psi^\dagger)\bigg(i\hbar\dot\psi-\frac12\frac{\delta 
h}{\delta\psi}\bigg)=0
\,,\label{ProjSchr1}
\eeq
along with the condition
\beq
\operatorname{Im}\left\langle\psi\bigg|\frac{\delta h}{\delta\psi}\right\rangle=0
\,,
\label{CompCond1}
\eeq
which implies that $h(\psi)$ is $U(1)-$invariant, since for a phase shift $\delta h = \langle i\psi ,{\delta h}/{\delta\psi}\rangle$.

Then, Noether's theorem for the $U(1)$ symmetry of the constrained Lagrangian in \eqref{VarPrinc2} again implies conservation of $\|\psi\|^2=\langle\psi,\psi\rangle$, since the Lagrangian is invariant under infinitesimal phase shifts. Consequently, the constraint $\|\psi\|^2-1=0$ is satisfied and we may write, simply,
\beq
i\hbar\dot\psi=\frac12\frac{\delta h}{\delta \psi}
\,,
\label{HamEq1}
\eeq
which is the Hamiltonian form of the class of Schr\"odinger equations. 

The Hamiltonian structure for the class of Schr\"odinger equations is encoded in the following symplectic form on $\mathscr{H}$:
$
\omega(\psi_1,\psi_2)=2\hbar\operatorname{Im}\langle\psi_1|\psi_2\rangle=2\hbar\langle i\psi_1, 
\psi_2\rangle
$.
In turn, this symplectic structure leads to the Poisson bracket given by
\[
\{f,g\}(\psi)=\frac1{2\hbar}\operatorname{Im}\left\langle\frac{\delta f}{\delta \psi}\bigg|\frac{\delta g}{\delta \psi}\right\rangle 
= \left\langle \frac{\delta f}{\delta \psi},-\frac{i}{2\hbar}\frac{\delta g}{\delta \psi} \right\rangle 
.\] 
This Poisson bracket then yields the corresponding Hamiltonian equation \eqref{HamEq1} via the expected relation $\dot{f}=\{f,h\}$.

Both the DF variational principle and the Hamiltonian structure presented above will be used again and again throughout this paper to illuminate the geometric features of current models in nonadiabatic molecular dynamics. The next section will review the geometric setting of Bohm's quantum hydrodynamics in terms of its Hamiltonian structure. In particular, the next section will show that the Hamiltonian functional $\langle\psi|\widehat{H}\psi\rangle$ \textit{{collectivizes}}, in the sense of Guillemin and Sternberg \cite{Guillemin,GuSt2013,LiMarle2012,marle2014,Marmo1985}, through the momentum maps leading from Schr\"odinger's equation to quantum hydrodynamics. 
A (left) Hamiltonian  action of a Lie group $G$ on a symplectic manifold $(M,\omega)$ induces the {\em momentum map} $J:M\to\mathfrak{g}^*$, where $\mathfrak{g}^*$ is the dual space to the Lie algebra $\mathfrak{g}$ of $G$. In the special case when $M$ is a symplectic vector space (so that $M=V$), then the momentum map is defined by
\beq
\langle J(x),\xi\rangle=\frac12\omega(\xi(x),x)
\,\ \qquad\forall x\in V\,,\quad\forall \xi\in \mathfrak{g}
\,,
\label{symp-momap}
\eeq
where $\xi(x)$ denotes the infinitesimal generator associated to the linear $G-$action on $V$. 
For example, if $G=SO(3)$, the momentum map $J(\bq,\bp)$ evaluates the angular momentum at each point $(\bq,\bp)\in \Bbb{R}^6$. 
In more generality, if $M$ is replaced by a Poisson manifold (so that $M=P$) with Poisson bracket $\{\cdot,\cdot\}_P$, the momentum map (if it exists) is defined as
\beq
\{F,\langle J(x),\xi\rangle\}_P=\xi[F]
\,\ \qquad\forall x\in P\,,\quad\forall \xi\in \mathfrak{g}\,,\quad\forall F\in C^\infty(P)
\,.
\label{Poisson-momap}
\eeq
Any function $h$ on $\mathfrak{g}^*$ then gives rise to a function $H = h\circ J$ on $M$ which is a {\em collective Hamiltonian} associated to the group action $G$. 
Symplectic momentum maps are Poisson. That is, for smooth functions $f$ and $h$, we have
$
\big\{F, H\big\} = \big\{f\circ J, h\circ J\big\} = \big\{f , h \big\}\circ J
$. 
This relation defines the {\em Lie-Poisson bracket} on $\mathfrak{g}^*$, given in terms of the adjoint action of the Lie algebra on itself, ${\rm ad}: \mathfrak{g}\times\mathfrak{g}\to \mathfrak{g}$, denoted as ${\rm ad}_\xi\zeta = [\xi,\zeta]$ for any Lie algebra elements $\xi,\zeta\in \mathfrak{g}$. Upon denoting the pairing by $\langle \,\cdot\,,\,\cdot\,\rangle_\mathfrak{g}: \mathfrak{g}^*\times \mathfrak{g}\to \bbR$, the Lie-Poisson bracket on $\mathfrak{g}^*$ reads as 
\cite{MaRa2013} 
\beq
\big\{ f(J)\,,\,h(J) \big\} 
:= \left\langle J\,,\, \left[ \frac{\partial f}{\partial J}\,,\,\frac{\partial h}{\partial J}\right] \right\rangle_\mathfrak{g}
=
-\left\langle J\,,\, {\rm ad}_{\partial h/\partial J}\frac{\partial f}{\partial J} \right\rangle_\mathfrak{g}
.\label{LPB-def}
\eeq
Momentum maps are ubiquitous in quantum mechanics. For example, the projection operator $J(\psi)=-i\hbar\psi\psi^\dagger$ is a momentum map for the left action of the unitary group $\mathcal{U}(\mathscr{H})$ on the quantum state space $\mathscr{H}$. Other important examples are given by quantum expectation values \cite{BLTr2016} and covariance matrices of  Gaussian wavepackets \cite{OhTr2017}.

\section{Quantum hydrodynamics}

This section illustrates the geometry of the hydrodynamic setting of quantum mechanics, which has its foundations in the Madelung transform \cite{Madelung1,Madelung2}. After reviewing the geometry of half-densities and their momentum maps, the latter are exploited to derive the quantum hydrodynamic (QHD) equations. 

\subsection{Half-densities and momentum maps }\label{1.3}
It is known \cite{Fusca,BatesWeinstein1997lectures} {that {wave function}s $\psi(\bx)$} in $\mathscr{H}=L^2(\Bbb{R}^3)$ can be regarded as half-densities, i.e., tensor fields $\psi\in \operatorname{Den}^{1/2}(\Bbb{R}^3)$ such that $|\psi|^2\in\operatorname{Den}(\Bbb{R}^3)$. More generally, if $\psi_1,\psi_2\in\operatorname{Den}^{1/2}(\Bbb{R}^3)$, then $\operatorname{Re}(\psi_1^*\psi_2)\in\operatorname{Den}(\Bbb{R}^3)$. The space $\operatorname{Den}^{1/2}(\Bbb{R}^3)$ is acted on by the diffeomorphism group $\operatorname{Diff}(\Bbb{R}^3)$ with the left action $\Phi:\operatorname{Diff}(\Bbb{R}^3)\times\operatorname{Den}^{1/2}(\Bbb{R}^3)\to\operatorname{Den}^{1/2}(\Bbb{R}^3)$ given by
\beq
\Phi(\eta,\psi) =:\Phi_\eta(\psi) = \frac{\psi\circ\eta^{-1}}{\sqrt{\operatorname{det}\nabla{\bbeta}^T}}
\,,
\label{half-dens-action}
\eeq
{where $\circ$ denotes composition of functions and ${\operatorname{det}\nabla\bbeta^T}$ denotes the Jacobian of the smooth invertible map, $\eta$, acting on coordinates $\bx\in\Bbb{R}^3$ as $\eta:\bx\mapsto\bbeta(\bx)\in\Bbb{R}^3 $.
The notation in \eqref{half-dens-action} defines} the mapping $\Phi_\eta:\operatorname{Den}^{1/2}(\Bbb{R}^3)\to\operatorname{Den}^{1/2}(\Bbb{R}^3)$ that is naturally induced by the group action {$\Phi(\eta,\psi)$ of the diffeomorphism $\eta\in\operatorname{Diff}(\Bbb{R}^3)$ on the half-density $\psi\in\operatorname{Den}^{1/2}(\Bbb{R}^3)$. Indeed, the left action in \eqref{half-dens-action} of diffeomorphisms on half-densities can be thought of as defining the push-forward of a half-density by a diffeomorphism.}

Upon using the anticommutator notation $\{A,B\}_+:=AB+BA$, the corresponding infinitesimal generator is given by
\beq
u_{\,\operatorname{Den}^{1/2}}(\psi)
=-\frac{i}{2}\hbar^{-1}\big\{\widehat{u}^k,\widehat{{P}}_k\big\}_+\psi
= -\bu\cdot\nabla\psi-\frac12(\nabla\cdot\bu)\psi
\,,
\label{Diff-action}
\eeq
where $\bu(\boldsymbol{x})\in \mathfrak{X}(\Bbb{R}^3)$ is a smooth vector field on $\Bbb{R}^3$, $\widehat{u}^{\,k\!}$ denotes the multiplicative operator associated to $u^k(\boldsymbol{x})$, and we recall that $\widehat{P}_k=-i\hbar\partial_{k}$ denotes the momentum operator. 

The  equivariant momentum map $\boldsymbol{J}:\operatorname{Den}^{1/2}(\Bbb{R}^3)\to\mathfrak{X}^*(\Bbb{R}^3)$ for the left action \eqref{Diff-action} is found as in \cite{Fusca,KhMiMo2017} from the standard definition \eqref{symp-momap}, that is $
\langle{\boldsymbol{J}}(\psi),\bu\rangle=\,\hbar\langle iu_{\,\operatorname{Den}^{1/2}}(\psi),\psi\rangle
$.
Here we have identified the Hilbert space as $\mathscr{H}=L^2(\Bbb{R}^3)=\operatorname{Den}^{1/2}(\Bbb{R}^3)$. 

Consequently, the space $\operatorname{Den}^{1/2}(\Bbb{R}^3)$ inherits the standard symplectic form on $L^2(\Bbb{R}^3)$. As a result, we have the 1-form density,
\beq
{\boldsymbol{J}}(\psi)=\operatorname{Re}(\psi^*\widehat{P}\psi)=\hbar\operatorname{Im}(\psi^*\nabla\psi)
=\hbar D\nabla{\theta}
\,,
\label{Jmomap-Clebsch}
\eeq
where the last equality follows from writing the wave function $\psi$ in polar form,
$
\psi=\sqrt{D}e^{i{\theta}}
$.
The momentum map $\bJ(\psi)$ coincides  (up to the  mass factor, $m$) with the well-known probability current from quantum mechanics. 
{It is also well known }that the physical Hamiltonian operator $\widehat{H}=-\hbar^2\Delta/2m+V(\boldsymbol{x})$ transforms the total energy into the form
\beq
\langle\psi|\widehat{H}\psi\rangle=\int\!\left[\frac1{2m}\frac{|\bJ(\psi)|^2}{|\psi|^2}+\frac{\hbar^2}{8m}\frac{(\nabla|\psi|^2)^2}{|\psi|^2}+|\psi|^2V(\boldsymbol{x})\right]\de^3x
= h(\bJ(\psi),|\psi|^2)
\,,
\label{collective1}
\eeq
as can be verified by a direct calculation. Here, the quantity $D=|\psi|^2$ arises as another momentum map which is associated to the action of (local) phase transformations $\psi(\boldsymbol{x})\mapsto e^{i\varphi(\boldsymbol{x})}\psi(\boldsymbol{x})$. Indeed, this action on the phase of the wave function has the momentum map $\hbar|\psi|^2=\hbar D$.

Relation \eqref{collective1} shows that the Hamiltonian functional $\langle\psi|\widehat{H}\psi\rangle$ \textit{collectivizes}, in the sense of Guillemin and Sternberg \cite{Guillemin,GuSt2013}, through the momentum maps $\bJ(\psi)$ and $|\psi|^2$. That is, the Hamiltonian in \eqref{collective1} may be expressed solely in terms of the collective variables $\boldsymbol\mu$ and $D$, given by
$
\langle\psi|\widehat{H}\psi\rangle=h(\boldsymbol\mu,D)$
with 
$\boldsymbol\mu=\bJ(\psi)$
and 
$(\bsmu,D)\in(\mathfrak{X}(\Bbb{R}^3)\,\circledS\, C^\infty(\Bbb{R}^3))^*$.
 {\color{black}Here,  $\mathfrak{X}(\Bbb{R}^3)\,\circledS\, C^\infty(\Bbb{R}^3)$ denotes the semidirect-product Lie algebra of the  semidirect-product Lie group $\operatorname{Diff}(\Bbb{R}^3)\,\circledS \,C^\infty(\Bbb{R}^3,S^1)$, whose elements $(\eta,\varphi)$ act from the left on the space $\operatorname{Den}^{1/2}(\Bbb{R}^3)$ of half-densities as 
\beq
\psi \mapsto \frac{1}{\sqrt{\operatorname{det}\nabla\bbeta^T}}\, ((e^{-i\varphi}\psi)\circ\eta^{-1})
\,.
\label{half-dens-action+phase}
\eeq
This formula extends the action in \eqref{half-dens-action} to include a local phase shift. }

The important feature here is that, under the {collectivization} $(\bJ(\psi),|\psi|^2)\to(\boldsymbol\mu,D)$, the Hamiltonian $h(\boldsymbol\mu,D)$ given by \eqref{collective1} belongs to a widely studied class of Hamiltonians possessing the Lie-Poisson bracket structure. This structure has been derived from the Euler-Poincar\'e formulation of ideal classical continuum dynamics with advected quantities in \cite{HoMaRa1998}. In particular, upon defining the velocity vector field {via the reduced Legendre transform},
$
\bu={\delta h}/{\delta \boldsymbol\mu}={m^{-1}}\boldsymbol\mu/D
$,
the Euler-Poincar\'e Lagrangian associated to the Hamiltonian \eqref{collective1} reads
\beq
\ell(\bu,D)=\int\!\left[\frac{mD}{2}|\bu|^2-\frac{\hbar^2}{8m}\frac{|\nabla D|^2}{D}-{D}V(\bsx)\right]\de^3x
\,.
\label{hydro-Lagr}
\eeq
{\color{black}Throughout this paper, we shall denote Euler-Poincar\'e Lagrangians by $\ell$ and ordinary Lagrangians by $L$. In \eqref{hydro-Lagr},  the velocity vector field is given by $\bu=\dot\eta\circ\eta^{-1}\in \mathfrak{X}(\Bbb{R}^3)$, while the Eulerian density $D$ is defined as
\beq
 \label{defns}
 D(\bx,t)=\eta_*D_0:=\int \!D_0(\bx_0)\,\delta(\bx-{\boldsymbol\eta}(\bx_0,t))\,\de^3x_0\in\operatorname{Den}(\Bbb{R}^3)\,,
 \eeq
for a reference density, $D_0=D_0(\bx)\,\de^3x$}. In the last definition, the symbol $\eta_*$ denotes the operation of {\em push-forward} by the map $\eta\in \operatorname{Diff}(\Bbb{R}^3)$; {so, $\eta_* D_0$ denotes the push-forward of the reference density, $D_0$ by the map $\eta$}. Push-forward by the smooth flow $\eta$ is called {\em advection} in hydrodynamics. {In this context, the Lagrangian particle path of a fluid parcel is given
by the smooth, invertible, time-dependent map, $\eta_t:\Bbb{R}^3\to\Bbb{R}^3$, as follows,
$
\eta_{t}\boldsymbol{x}_0=\boldsymbol{\eta}(\boldsymbol{x}_0,t) \in\mathbb{R}^{3} 
$
{for initial reference position}  
$\eta_{0}\boldsymbol{x}_0=\boldsymbol{\eta}(\boldsymbol{x}_0,0)=\boldsymbol{x}_0$.
After this definition, there should be no confusion between $\eta_{t}\in {\rm Diff}(\mathbb{R}^3)$
and $\eta_{t}\boldsymbol{x}_0 = \boldsymbol{\eta}(\boldsymbol{x}_0,t) \in\mathbb{R}^{3}$. The subscript $t$ is omitted in most of this paper, for simplicity of notation.}

Now, taking variations in Hamilton's principle $\delta \int_{t_1}^{t_2} \ell(\bu,D)\,\text{d}t=0$ for the reduced Lagrangian   \eqref{hydro-Lagr} yields the following quantum hydrodynamics (QHD) equations,
\begin{equation}
  \frac{\partial \bu}{\partial t} + (\bu\cdot\nabla)\bu = -\frac{1}{m}\nabla(V+V_Q)
\,,\qquad\qquad\ 
  \frac{\partial D}{\partial t} + \text{div}(D\bu) = 0\,.
  \label{u-eqn1}
\end{equation}
Here,  the {\it quantum potential}
\beq
V_Q:= -\,\frac{\hbar^2}{2m}\frac{\Delta \sqrt{D}}{\sqrt{D}}
\label{qPotential-def}
\eeq
arises from taking variations of the middle term of the reduced Lagrangian in \eqref{hydro-Lagr}, 
which can be rearranged as {$|\nabla D|^2/D=4|\nabla \sqrt{D}|^2$}. 

\begin{remark}[Effects of the quantum potential]
The appearance of the amplitude of the wave function in the denominator of the quantum potential in \eqref{qPotential-def} implies that its effects do not necessarily fall off with distance. That is, the effects of the quantum potential need not decrease, as the amplitude of the wave function decreases. Moreover, the middle term in \eqref{hydro-Lagr} is known as the Fisher-Rao norm, which is well-known in information theory. For further discussion of the information geometry in quantum mechanics, see, e.g., \cite{BrHu1998}.
\end{remark}

Equations \eqref{u-eqn1} follow from Hamilton's principle  for the reduced (collective) Lagrangian  \eqref{hydro-Lagr}, upon using the following constrained variations from the Euler-Poincar\'e theory of ideal fluids with advected quantities, derived in \cite{HoMaRa1998},
\beq
\delta\bu=\delta(\dot\eta\circ\eta^{-1})=\partial_t\bw+(\bu\cdot\nabla)\bw-(\bw\cdot\nabla)\bu
\,,\qquad
\delta D=\delta (\eta_*D_0) =-\operatorname{div}(D\bw)
\,.
\label{EPvar}
\eeq
Here, the arbitrary vector field $\bw=\delta\eta\circ\eta^{-1}\in\mathfrak{X}(\Bbb{R}^3)$ vanishes at the endpoints in time. The density $D$ is an advected quantity, satisfying the mass transport equation in \eqref{u-eqn1}.

\subsection{Bohmian trajectories, Lagrangian paths \& Newton's Law\label{Bohmian-paths}}

In quantum hydrodynamics, the role of the Lagrangian path $\eta\in\operatorname{Diff}(\Bbb{R}^3)$ is of paramount importance. Namely, it plays the role of a {\it hidden variable} in the Bohmian interpretation of quantum dynamics \cite{Bohm}. Indeed, in this framework the path ${\color{black}\eta}$ is the fundamental dynamical variable, while the {wave function} is {simply} transported in time along the Lagrangian motion of $\bbeta(\bx_0,t)$, which in turn satisfies
\beq
\dot{\bbeta}(\bx_0,t)=\bu(\bbeta(\bx_0,t),t)\,.
\label{EPvelocity}
\eeq
This relation defines the so-called \emph{Bohmian trajectory}, which is precisely the Lagrangian fluid path of the hydrodynamic picture!


At this point, it is important to {emphasize} that the (infinite-dimensional) Bohmian trajectories {$\bbeta(\bx_0,t)$} are completely different from the point particle trajectories $\bq(t)$ (finite-dimensional), which arise when the quantum dispersion is neglected. In order to clarify this point, it is convenient to rewrite the Lagrangian \eqref{hydro-Lagr} in terms of the Bohmian trajectory by using \eqref{defns} and \eqref{EPvelocity}. We have
 \beq
L(\bbeta,\dot\bbeta)=\int\!\left[\frac{mD_0}{2}{|\dot\bbeta|^2}-D_0(\bx_0)\Big(V_Q(\bbeta(\bx_0,t),t)+V(\bbeta(\bx_0,t))\Big)\right]\de^3x_0
\,,
\label{hydro-Lagr1}
\eeq
where the quantum potential is  written in terms of $\bbeta$ as
\[
V_Q(\bx,t)=-\,\frac{\hbar^2}{2m}\sqrt{
\frac{
\operatorname{det}\nabla\bbeta(\bx,t)^T
}
{D_0(\bbeta^{-1}(\bx,t))}}
\,
\Delta\, \sqrt{\frac{D_0(\bbeta^{-1}(\bx,t))}{\operatorname{det}\nabla\bbeta(\bx,t)^T}}
\,.
\]
The dynamics of the Bohmian trajectory $\eta$ is given by the Euler-Lagrange equation \cite{Wyatt}
$
mD_0\ddot\bbeta=-D_0\nabla_{\bbeta}( V_Q(\bbeta,t)+V(\bbeta))
$.
{We emphasize that the dynamics of the Bohmian trajectory $\eta$  is not equivalent to point particle dynamics.} In principle, the latter could be obtained by setting a point-like initial density of the type $D_0(\bx_0)=\delta(\bx_0-\bq_0)$ and then integrating the Euler-Lagrange equation over $D_0$. However, this type of initial condition  is not allowed by the structure of the quantum potential. For this reason, asymptotic semiclassical methods are required to properly derive the effects of the quantum potential in a weak limit as $\hbar^2\to0$. For more details, see, e.g., \cite{JinLi2003}.

Nonetheless, the Newtonian limit neglects the order $O(\hbar^2)$ quantum dispersion term in the Lagrangian \eqref{hydro-Lagr} (or, equivalently, \eqref{hydro-Lagr1})
and varies the remainder. The resulting equation for the Bohmian trajectory becomes
$
D_0\big(m\ddot\bbeta+\nabla_{\bbeta}V(\bbeta)\big)=0
$.
It is clear that the point-particle initial condition $D_0(\bx_0)=\delta(\bx_0-\bq_0)$ is now allowed and thus denoting $\bq(t)=\bbeta(\bq_0,t)$ and integrating over space yields Newton's Law
$m\ddot\bq+\nabla V(\bq)=0$.

In the Eulerian picture, one proceeds analogously by discarding the $O(\hbar^2)$ in the Lagrangian \eqref{hydro-Lagr}, so that the equations of motion \eqref{u-eqn1} restrict to
\beq
D(\partial_t\bu+\bu\cdot\nabla\bu)=-m^{-1}D\nabla V
\,,\qquad\quad
\partial_t D+\operatorname{div}(D\bu)=0
\,.
\label{Restricted-qHcH-eqns}
\eeq
Then, one considers the relations \eqref{defns} and
\eqref{EPvelocity} between Lagrangian and Eulerian quantities.  
 We observe that the initial particle-type initial density $D_0(\bx_0)=\delta(\bx_0-\bq_0)$ yields the Eulerian relation $D(\bx,t)=\delta(\bx-\bq(t))$ with $\dot{\bq}=\bu(\bq,t)$. Integrating \eqref{Restricted-qHcH-eqns} over space again recovers Newton's Law for $\bq(t)$.


A common alternative method to derive Newton's Law exploits the analogy with the Hamilton-Jacobi equation of classical mechanics. 
Since $\bu=m^{-1}\boldsymbol\mu/D=\hbar \nabla {\theta}$ according to the momentum map relation for the collective variable $\bJ(\psi)$ in \eqref{Jmomap-Clebsch}, the first of these restricted QHD equations happens to recover the Hamilton-Jacobi equation for $S=\hbar{\theta}$, as for geometrical optics with the classical Hamiltonian $H(\bq,\bp)=|\bp|^2/2M+V(\bq)$. This is not necessarily convenient for a fluids interpretation, though, because solutions of Hamilton--Jacobi equations may become singular (e.g., form caustics) even for smooth initial data.


\begin{remark}[Regularization of an ``ultraviolet catastrophe'' for $\hbar^2\to0$]\label{ultra}
The Newtonian limit of QHD \eqref{u-eqn1}, obtained by simply neglecting the contribution to the Euler-Poincar\'e equations from the quantum potential in \eqref{qPotential-def} turns out to be problematic. In particular, because the {potential (which plays the role of a pressure term)} is assumed to be independent of time, the Newtonian limit system \eqref{Restricted-qHcH-eqns} is not strictly hyperbolic. This observation is a well known signal, \cite{LaLi2013}, that the solution behaviour in the classical limit $\hbar^2\to0$ can become singular, as $\hbar^2$ multiplies the highest spatial derivative. This is especially clear when the {wave function} is written in the usual WKB form, as $\psi=\sqrt{D}\exp(iS/\hbar)$, where $S$ is the action integral for the Schr\"odinger equation.
Indeed, in the $\hbar^2\to0$ limit, the gradient of the quantum potential produces highly oscillatory spatial behaviour. See, e.g., \cite{GaMa1997,JinLi2003} and references therein for discussions of the weak convergence of the rapidly oscillatory solutions obtained in passing the WKB description of the Schr\"odinger equation to the classical limit as $\hbar^2\to0$. 
As indicated in \cite{JinLi2003}, one convenient way to carry out the limit $\hbar^2\to0$ is to apply the Wigner transform of the wave function \cite{Wigner1932,ZaFaCu2005}. The real-valued Wigner function then acts as the quantum-equivalent of the phase-space distribution function in classical mechanics; although the quantum version introduces mathematically technical features, such as Moyal operator brackets, instead of Poisson brackets. One could also treat the rapid oscillations as being stochastic and use probability theory to obtain the expected solution as a classical limit, \cite{Streater2000}. 
Thus, in retrospect, one can appreciate the role of non-zero $\hbar^2$ in the quantum potential \eqref{qPotential-def} in equations \eqref{u-eqn1} as a dispersive {regularization} of what would otherwise have led to a type of {\em ultraviolet catastrophe} \cite{Co1977} for the restricted (Hamilton-Jacobi) QHD in \eqref{Restricted-qHcH-eqns}, as the solutions of the restricted QHD equations form caustics. 
\end{remark}

\subsection{Lie-Poisson structure of quantum hydrodynamics}\label{LPHam-QHE}

In terms of the variables $(\boldsymbol\mu,D)$, the collective Hamiltonian in \eqref{collective1} for the QHD equations in \eqref{u-eqn1} reads
\begin{align}
h(\boldsymbol\mu,D)=
\int\left(\frac{1}{2m}\frac{|\boldsymbol\mu|^2}{D} + \frac{\hbar^2}{8m}\frac{|\nabla D|^2}{D}
+ DV(\boldsymbol{x})\right)\,\text{d}^3x  
\,. \label{collective2}
  \end{align}
In these variables, the QHD equations can be {written} in Hamiltonian form, with a Lie-Poisson bracket written symbolically as, see, e.g., \cite{HoMaRa1998}
%
\begin{equation} 
\frac{\partial}{\partial t}
    \begin{bmatrix}
    \boldsymbol{\mu} \\ D 
    \end{bmatrix}
=
\left\{
    \begin{bmatrix}
    \boldsymbol{\mu} \\ D 
    \end{bmatrix}
    \,,\,
    h(\boldsymbol{\mu},D)
\right\}
=
    -\,{\cal C} 
   \begin{bmatrix}
   \delta h/\delta \boldsymbol{\mu} \\
   \delta h/\delta D    
    \end{bmatrix}    
= -\, 
\begin{bmatrix}
   {\rm ad}^\ast_\Box\,\boldsymbol{\mu} &
   \Box\diamond D \\
   \pounds_\Box\,D & 0 
    \end{bmatrix}
   \begin{bmatrix}
   \delta h/\delta \boldsymbol{\mu} \\
   \delta h/\delta D    
    \end{bmatrix}    \,,
\label{LP-Ham-1}
\end{equation}
in which each box $\Box$ in \eqref{LP-Ham-1} indicates where to substitute elements of the last column 
of variational derivatives of the Hamiltonian  in the matrix multiplication. Here ${\rm ad}^\ast$ denotes the coadjoint
action of the Lie algebra $\mathfrak{X}(\Bbb{R}^3)$ on its dual, the 1-form densities $\mathfrak{X}(\Bbb{R}^3)^*=\Lambda^1(\Bbb{R}^3)\otimes \operatorname{Den}(\Bbb{R}^3)$. The coadjoint action ${\rm ad}^*: \mathfrak{g}\times\mathfrak{g}^*\to \mathfrak{g}^*$ in \eqref{LP-Ham-1} is dual to the adjoint action ${\rm ad}: \mathfrak{g}\times\mathfrak{g}\to \mathfrak{g}$, via the pairing $\langle \,\cdot\,,\,\cdot\,\rangle_\mathfrak{g}: \mathfrak{g}^*\times \mathfrak{g}\to \bbR$ in which the Lie-Poisson bracket in equation \eqref{LPB-def} was defined, see, e.g., \cite{HoScSt2009,MaRa2013},
$
\big\langle {\rm ad}^\ast_{\partial h/\partial \boldsymbol{\mu}}\,\boldsymbol{\mu} \,,\,
{\partial f}/{\partial \boldsymbol{\mu}} \big\rangle_\mathfrak{g}
:=
\left\langle \boldsymbol{\mu}\,,\, {\rm ad}_{\partial h/\partial \boldsymbol{\mu}}({\partial f}/{\partial \boldsymbol{\mu}}) \right\rangle_\mathfrak{g}
$.
The symbol $\pounds_\bu$ in \eqref{LP-Ham-1} denotes Lie derivative with respect to the vector field $\bu=\dot{\eta}\circ\eta^{-1}\in\mathfrak{X}(\Bbb{R}^3)$. For example, the corresponding Lie derivative of the density {$D(\bx,t)\text{d}^3x$} is given by
\beq
\pounds_\bu (D(\bx,t){\text{d}^3x}) = {\frac{\text{d}}{\text{d}t}}\bigg|_{t=0}\big(D(\bbeta(\bx,t),t)\,{\text{d}^3\eta(\bx,t)}\big)
=
{\rm div} \big( \bu D(\bx,t)\big){\text{d}^3 x}
\,,
\label{Lie-der}
\eeq
for the Lagrangian path, $\bbeta(\bx,t)$ such that $\bbeta(\bx,0)=\bx$. Finally, the diamond operation $(\diamond)$ is defined for right action of $\Phi_{\eta(t)}$ as \cite{HoScSt2009}
\beq
\left\langle \frac{\delta h}{\delta a} \diamond a \,,\,\xi \right\rangle_\mathfrak{g}
:=
\left\langle \frac{\delta h}{\delta a}\,,\, - \pounds_\xi a  \right\rangle_{V^*\times V}
,
\label{diamond-def}
\eeq
in the $L^2(\Bbb{R}^3)$ pairing $\langle \,\cdot\,,\,\cdot\,\rangle_{V^*\times V}: V^*\times V \to \bbR$ for elements of the tensor space $a\in V$ and on its dual $\delta h/\delta a\in V^*$. In the example of the advected density, we have $D\in \operatorname{Den}(\Bbb{R}^3)$.

The corresponding notation is defined explicitly for QHD {\em in components} by 
\begin{align}
\begin{split}
({\rm ad}^\ast_{\delta h/\delta \boldsymbol{\mu}}\,\boldsymbol{\mu})_i 
&= 
(\partial_j \mu_i + \mu_j \partial_i)\frac{\delta h}{\delta \mu_j} 
\in\Lambda^1(\Bbb{R}^3)\otimes \operatorname{Den}(\Bbb{R}^3)\,,
\\
\pounds_{\delta h/\delta \boldsymbol{\mu}}\,D &:= {\rm div} \left( D \frac{\delta h}{\delta \boldsymbol{\mu}} \right)
\in \operatorname{Den}(\Bbb{R}^3)\,,
\\
\frac{\delta h}{\delta D}\diamond D &:= D \nabla \frac{\delta h}{\delta D} 
\in \Lambda^1(\Bbb{R}^3)\otimes \operatorname{Den}(\Bbb{R}^3)\,.
\end{split}
\label{LPB-notation}
\end{align}

The Hamiltonian operator ${\cal C}$ in \eqref{LP-Ham-1} is linear in the dynamical variables $(\boldsymbol\mu,D)$.
The corresponding Lie-Poisson bracket is given explicitly by
{\begin{align}
\begin{split}
\{f\,,\,h\}(\mu,D) 
&= - \int {\rm Tr} \left(\left(\frac{\delta f}{\delta (\boldsymbol{\mu},D)}\right)^T{\cal C}\,\frac{\delta h}{\delta (\boldsymbol{\mu},D)}\right)\,\text{d}^3x
\\&= - \int \mu_j \left( \frac{\delta f}{\delta \boldsymbol{\mu}} \cdot \nabla  \frac{\delta h}{\delta \mu_j}
- \frac{\delta h}{\delta \boldsymbol{\mu}} \cdot \nabla  \frac{\delta f}{\delta \mu_j} \right)
\\& \hspace{15mm}
+ D \left( \frac{\delta f}{\delta \boldsymbol{\mu}} \cdot \nabla  \frac{\delta h}{\delta D}
- \frac{\delta h}{\delta \boldsymbol{\mu}} \cdot \nabla  \frac{\delta f}{\delta D} \right)\,\text{d}^3x
\,.
\end{split}
\label{LPB}
\end{align}}
This is the Lie-Poisson bracket dual to the semidirect product Lie algebra $\mathfrak{X}(\mathbb{R}^3)\,\circledS\, C^\infty(\mathbb{R}^3)$, with dual coordinates $\boldsymbol{\mu}\in \mathfrak{X}^*(\mathbb{R}^3)=\Lambda^1(\mathbb{R}^3)\otimes {\rm Den}(\mathbb{R}^3)$ (1-form densities) and $D\in {\rm Den}(\mathbb{R}^3)$. For further discussions of {Lie-Poisson brackets,} see, e.g., \cite{MaRa2013,HoScSt2009} and references therein.

\subsection{Regularized QHD and Bohmion solutions}\label{QHE-Bohmions}

In remark \ref{ultra}, we have {emphasized} that the order $O(\hbar^2)$ term in the QHD Hamiltonian  \eqref{collective2} may be regarded as a dispersive {regularization} of the ``ultraviolet catastrophe'' which occurs in the quantum fluid Hamiltonians as $\hbar^2\to0$. The order $O(\hbar^2)$ term is an energy penalty for high gradients $|\nabla D|^2/D=4|\nabla \sqrt{D}|^2$ that yields only a weak classical limit as $\hbar^2\to0$ \cite{JinLi2003}.

In this section we proceed by {regularizing} the Lagrangian or Hamiltonian to allow for single-particle solutions. As we have observed in Section \ref{Bohmian-paths}, the $O(\hbar^2)-$terms in QHD prevent the existence of particle-like solutions so that Bohmian trajectories can only be identified with Lagrangian paths  following the characteristic curves of the Eulerian fluid velocity. Thus, the $O(\hbar^2)-$terms in QHD must be treated with particular attention. Instead of adopting semiclassical methods to take the limit {$\hbar^2\to 0$}, this section presents an alternative strategy consisting in {regularizing the} $O(\hbar^2)-$terms by a smoothening process.  More particularly, we shall discuss Bohmian trajectories which can be computed from {regularized} QHD Hamiltonians and Lagrangians, whose fluid variables have been spatially smoothed; so that their $\hbar^2\to0$ limit is no longer singular.

Depending on which terms are regularized, different particle motions may emerge. We present two regularization strategies. 
The first simply smoothens all the terms in the Hamiltonian, while the second only smoothens the $O(\hbar^2)-$terms. Although all the equations of motion derived here are Hamiltonian equations on a canonical phase space, they may or may not be in the  usual Newtonian form, depending on which {regularization} scheme is adopted. In particular, the first {regularization} scheme is adopted in the Hamiltonian framework to {regularize} the hydrodynamic momentum and density, while the second scheme is based on the variational approach following the standard Bohmian method.

\paragraph{Hamiltonian regularization.}
The first {regularization} introduces the mollified collective Hamiltonian for {\em {regularized} quantum hydrodynamics (RQHD)}. This is simply obtained by replacing the variables $\bsmu$ and $D$ in \eqref{collective2} by the corresponding spatially smoothed 
variables,
\beq
\bsmu\to \bsmubar = K*\bsmu = \int \!K(\bx,\bs)\bsmu (\bs) \,{\de}^3s
\quad\hbox{and}\quad
D\to \Dbar = K*D = \int \!K(\bx,\bs)D (\bs){\de}^3s\,,
\label{LPB-Hambar}
\eeq
where $K(\bx,\bs)$ is a positive definite, symmetric smoothing kernel which falls off at least exponentially in $|\bx-\bs|$.
For example, we may take the kernel $K(\bx,\bs)$ to be the Green's function of the Helmholtz operator ${(1-\alpha^2\Delta)}$, where $\alpha$  is a length scale and the limit $\alpha\to 0$  returns the original  hydrodynamic variables. This choice of smoothing kernel is also an energy penalty for high gradients. It promotes functions $(\bsmu,D)$ that are bounded in the $L^2$ norm to functions $(\bsmubar,\Dbar)$ that are bounded in the $H^1$ Sobolev norm.  Similarly, choosing $K(\bx,\bs)$ to be the Greens function for an integer power $p$ of the Helmholtz operator would smooth gradients to promote functions of $(\bsmubar,\Dbar)$ from being bounded in $L^2$, to being bounded in the Sobolev space, $H^p$. Yet another choice would be to take $K(\bx,\bs)$ to be a Gaussian.

Upon replacing the variables $(\boldsymbol\mu,D)$ in \eqref{collective2} for Hamiltonian $h(\boldsymbol\mu,D)$ by the spatially smoothed variables $(\bsmubar,\Dbar)$ as $h(\bsmubar,\Dbar)=h(K*\boldsymbol\mu,K*D)$, we find the following Hamiltonian equations for the original variables, $(\bsmu,D)$,
\begin{equation} 
\frac{\partial}{\partial t}
    \begin{bmatrix}
    \boldsymbol{\mu} \\ D 
    \end{bmatrix}
=
\left\{
    \begin{bmatrix}
    \boldsymbol{\mu} \\ D 
    \end{bmatrix}
    \,,\,
    h(\bar{\boldsymbol{\mu}},\Dbar)
\right\}
= -\, 
\begin{bmatrix}
   {\rm ad}^\ast_\Box\,\boldsymbol{\mu} &
   \Box\diamond D \\
   \pounds_\Box\,D & 0 
    \end{bmatrix}
   \begin{bmatrix}
   K*(\delta h/\delta \bar{\boldsymbol{\mu}}) \\
   K*(\delta h/\delta \Dbar)    
    \end{bmatrix}    \,.
\label{LP-Ham-2}
\end{equation}
Of course, the {regularized} quantum hydrodynamics (RQHD) equations arising after replacing the Hamiltonian in \eqref{collective2} by $h_\textit{\tiny RQHD}=h(\bsmubar,\Dbar)$ must take the same Lie-Poisson form as in equations \eqref{LP-Ham-1}, modified now to read
{\begin{align}
h_\textit{\tiny RQHD}(\bsmu,D) = h(\bsmubar,\Dbar)=
\int\left(\frac{1}{2m}\frac{|\bsmubar|^2}{\Dbar} + \frac{\hbar^2}{2m}\big|\nabla \sqrt{\Dbar}\,\big|^2
+ \Dbar V(x)\right)\,\text{d}^3x 
\,. \label{collective2-Bohm}
  \end{align}}

The variations of $ h_\textit{\tiny RQHD}(\bsmu,D)$ are given by the following $L^2$ functions for an appropriate choice of the kernel $K$, rewritten in the same form as for the unsmoothed variables,
\begin{align}
\frac{\delta h_\textit{\tiny RQHD}}{\delta \bsmu}
= K*\frac{\delta h}{\delta \bsmubar} &=   K*\left(\frac{\bsmubar}{m\Dbar}\right)
\,,\\
\frac{\delta h_\textit{\tiny RQHD}}{\delta D}
= K*\frac{\delta h}{\delta \Dbar} &= - K*\left(
\frac{|\bsmubar|^2}{2m\Dbar^2} + \frac{\hbar^2}{2m} \frac{\Delta \sqrt{\Dbar}}{\sqrt{\Dbar}} 
- V(x)
\right)
. \label{Bohmion-deltaHambar}
  \end{align}  
At this point, the advantage of having {regularized} by simply smoothing the variables in the Hamiltonian by $K*$ will emerge. Namely, while the physical meaning of the various expressions in the Hamiltonian has been preserved, the solutions in the original variables $(\bsmu,D)$ can now be {\em singular and finite dimensional} along the Lagrangian paths for the diffeomorphism, $\eta$ in Section \ref{Bohmian-paths}. 

Specifically, equations \eqref{LP-Ham-2} for $h_\textit{\tiny RQHD}=h(\bsmubar,\Dbar)$ admit Lagrangian paths as particle-like singular solutions for $(\boldsymbol{\mu},D)$, which we propose to call `Bohmions'. These are given  by the singular momentum map 
\begin{align}
\boldsymbol{\mu}(\bx,t) = \sum_{a=1}^{\cal N}  \boldsymbol{p}_a(t) \delta(\bx-\boldsymbol{q}_a(t))
\,,\qquad\qquad 
D(\bx,t) =   \sum_{a=1}^{\cal N} w_a \delta(\bx-\boldsymbol{q}_a(t))
\,, \label{Bohmion-sing-solns1D}
  \end{align}  
  with $\sum_a w_a =1$. 
The momentum map \eqref{Bohmion-sing-solns1D} was presented in \cite{HoTr2009} as the immediate extension of the singular momentum map represented by the first relation in \eqref{Bohmion-sing-solns1D} and  first discovered in \cite{HoMa2005}. This singular momentum map underlies the `peakon' singular solutions for the nonlinear wave/fluid equations in \cite{CaHo1993,HoONTr2009}. These previously investigated `peakon' singular solutions were obtained, respectively, by smoothing the momentum in the kinetic energy for either geodesic motion on Diff$(\bbR^3)$ in the case of \cite{CaHo1993}, and by smoothing both the momentum and the depth for shallow water dynamics in \cite{HoONTr2009}. {\color{black}Before proceeding further, let us emphasize that the formal process leading to the singular solutions \eqref{Bohmion-sing-solns1D} has provided a Lagrangian-particle representation of the dynamics. However, this does not imply that the variables $D = |\psi|^2$ and $\bmu=\hbar\operatorname{Im}(\psi^*\nabla\psi)$ will become delta functions. Nevertheless, as we shall see below, the dynamics emerging from the singular expressions \eqref{Bohmion-sing-solns1D} does reveal the Lagrangian-particle content of quantum hydrodynamics. Namely, the Bohmian `particles' are merely fluid labels following Lagrangian flow trajectories. As such, the Bohmion singular solutions in \eqref{Bohmion-sing-solns1D} represent flow lines in quantum hydrodynamics as virtual particles. Of course, this is not a radically new idea, since particle methods have a long and successful history in computational fluid dynamics. }

Upon restricting to consider smoothing kernels of the type $K(\bx,\bs)=K(\bx-\bs)$, substitution of the Bohmion singular solutions \eqref{Bohmion-sing-solns1D} into the Hamiltonian $h_\textit{\tiny RQHD}=h(\bar{\boldsymbol{\mu}},\Dbar)$ in \eqref{collective2-Bohm}, with regularized quantities
\begin{align}
\begin{split}
\bar{\boldsymbol{\mu}}(\bx,t) =&\  K*\boldsymbol{\mu} =  \sum_{a=1}^{\cal N} \boldsymbol{p}_a(t) K(\bx-\boldsymbol{q}_a(t))\,,
\\
\Dbar(\bx,t) = &\  K*D = \sum_{a=1}^{\cal N} w_a K(\bx-\boldsymbol{q}_a(t))
\,, 
\end{split}
\label{Bohmion-Ksolns1Dbar}
\end{align}
one evaluates the Bohmion Hamiltonian $h_\textit{\tiny RQHD}(\{\boldsymbol{q}_a\},\{\boldsymbol{p}_a\})$ in terms of its canonical phase space variables $(\{\boldsymbol{q}_a\},\{\boldsymbol{p}_a\})$, as
\begin{align}
\begin{split}
h_\textit{\tiny RQHD}(\{\boldsymbol{q}_a\},\{\boldsymbol{p}_a\}) 
&= \frac{1}{2m}
{\int \frac{\sum_{a,b} \boldsymbol{p}_a\cdot\boldsymbol{p}_bK(\boldsymbol{s}-\boldsymbol{q}_a)K(\boldsymbol{s}-\boldsymbol{q}_b)}{\sum_{c} w_cK(\boldsymbol{s}-\boldsymbol{q}_c)} \,d^3s }
\\&\quad 
+ \frac{\hbar^2}{8m}
{ \int \frac{\sum_{a,b} w_aw_b\nabla K(\boldsymbol{s}-\boldsymbol{q}_a)\cdot\nabla K(\boldsymbol{s}-\boldsymbol{q}_b)}{\sum_{c} w_cK(\boldsymbol{s}-\boldsymbol{q}_c)} \,d^3s }
\\&\quad + \sum_{a} w_a  \int K(\boldsymbol{s}-\boldsymbol{q}_a) V(s)  \,d^3s
\,,
\end{split}
\label{HamBohm-pq}
\end{align}
Then, according to equivariance of the momentum map \eqref{Bohmion-sing-solns1D} discovered in \cite{HoMa2005,HoTr2009}, the dynamics of $(\boldsymbol{q},\boldsymbol{p})$ satisfy the canonically conjugate Hamiltonian equations in phase space,
\begin{align}
\dot{\boldsymbol{q}}_a=  \frac{\delta h_\textit{\tiny RQHD}}{\delta \boldsymbol{p}_a} = \bu(\boldsymbol{q}_a(t),t)
\quad\hbox{and}\quad
\dot{\boldsymbol{p}}_a =  -\,\frac{\delta h_\textit{\tiny RQHD}}{\delta \boldsymbol{q}_a} \,.
\label{HamcanonBohmion-eqns2}
  \end{align}
Both the momentum term and the quantum term proportional to $\hbar^2$ in the canonical equations \eqref{HamcanonBohmion-eqns2} provide extensive, potentially long-range coupling among the singular particle-like Bohmion solutions, because of the presence of $\Dbar$ in the denominators of these terms in the Hamiltonian $h_\textit{\tiny RQHD}$ in \eqref{collective2-Bohm}. However, the limit $\hbar^2\to0$ in the canonical Bohmion equations \eqref{HamcanonBohmion-eqns2} is no longer singular. 

\paragraph{Lagrangian regularization.}
So far, the discussion has been entirely based on the Hamiltonian structure of QHD. In this section, we shall take an alternative  route: instead of {regularizing} all terms in the QHD Hamiltonian, we shall {regularize} only the $O(\hbar^2)-$terms in the QHD Lagrangian \eqref{hydro-Lagr}. As a first step, we perform the substitution $D\to\bar{D}$ in the $O(\hbar^2)-$term of the original QHD Lagrangian \eqref{hydro-Lagr}, which then becomes
\beq
\ell(\bu,D)=\int\!\left[\frac{mD}{2}|\bu|^2-\frac{\hbar^2}{8m}\frac{| \nabla\bar{D}|^2}{\bar{D}}-{D}V\right]\de^3 x
\,.
\label{hydro-RLagr}
\eeq
Then, upon following the arguments in the previous section, we set the initial condition 
\beq
D_0(\bx)=\sum_{a=1}^{\cal N} w_a\delta(\bx- \boldsymbol{q}_a{(0)})
\label{D-ini}
\eeq
(with $\sum_a w_a =1$)
and denote $\boldsymbol{q}_a(t)=\boldsymbol\eta(\boldsymbol{q}_a{(0)},t)$, where $\boldsymbol\eta({\bx},t)$ is the Lagrangian path such that $\dot{\boldsymbol\eta}(t)=\bu(\boldsymbol\eta(t),t)$. Then, upon recalling the Lagrange-to-Euler map in \eqref{defns}, the Eulerian density becomes
$
D(\bx,t)=\sum_aw_a \delta(\bx- \boldsymbol{q}_a(t))
$.
In turn, this expression can be inserted into the {regularized} QHD Lagrangian \eqref{hydro-RLagr}, which becomes
\begin{multline}
L(\{\boldsymbol{q}_a\},\{\dot{\boldsymbol{q}}_a\}) 
=\sum_{a} \bigg(
\frac{mw_a}2|\dot{\boldsymbol{q}}_a|^2 
\\{
- \frac{\hbar^2}{8m}\int \frac{\sum_{b} w_aw_b \nabla K(\boldsymbol{y}-\boldsymbol{q}_a)\cdot\nabla K(\boldsymbol{y}-\boldsymbol{q}_b)}{\sum_{c} w_c K(\boldsymbol{y}-\boldsymbol{q}_c)} \,d^3y}
- w_aV(\boldsymbol{q}_a)
\bigg)
\end{multline}
for which Hamilton's principle produces the Euler-Lagrange equation
\[
m\ddot{\boldsymbol{q}}_a=-\nabla V(\boldsymbol{q}_a)
-\frac{\hbar^2}{8m} {
 \frac{\partial}{\partial \boldsymbol{q}_a}\int \frac{\sum_{b} w_b \nabla K(\boldsymbol{y}-\boldsymbol{q}_a)\cdot\nabla K(\boldsymbol{y}-\boldsymbol{q}_b)}{\sum_{c} w_c K(\boldsymbol{y}-\boldsymbol{q}_c)} \,d^3y }
 \,.
 \]
{\color{black}In analogy to the arguments in the previous section, we emphasize that the formal process outlined here reveals a form of Newtonian dynamics, which is suitably modified by the regularized expression of the quantum potential. While the formal relation \eqref{D-ini} provides a particle description, it does not imply, for example, that a smooth initial probability distribution  $D=|\psi|^2$ would evolve to concentrate into a delta function. }

\subsection{Density operators and classical closures\label{densities}}

So far, the discussion has focused uniquely on wave functions for pure quantum states. However, mixed quantum states are a more general class of states that can be represented by a Hermitian, unit-trace, and positive-definite  integral operator $\rho$ satisfying the Liouville-von Neumann equation 
\beq\label{LvN}
i\hbar\,\partial_t\rho=[\widehat{H},\rho\,]
\,.
\eeq
Pure states are regarded as a special case of mixed states under the identification ${\rho}=\psi\psi^\dagger$, or equivalently $\rho(\bx,\bx')=\psi(\bx)\psi^*(\bx')$. Here, the notation $\rho(\bx,\bx')$ is used for the kernel ({\em matrix element}, in the physics literature) of the integral operator $\rho$. In more generality, for an arbitrary sequence $\{\psi_n(\bx)\}$ of $N$ square-integrable functions, the {\em density operator} is given by 
\begin{equation}\label{rhomixture}
\rho(\bx,\bx')=\sum_{n=1}^N w_n\,\psi_n(\bx)\psi_n^*(\bx')
\end{equation}
with $\sum_{n} w_n=1$. For the momentum map aspects of quantum mixed states, see \cite{Montgomery,Tronci2018}. Equation \eqref{LvN} is of Lie-Poisson type, with the bracket structure
\[
\{f,h\}(\rho\,)=-i\hbar^{-1}\operatorname{Tr}\left(\rho\left[\frac{\delta f}{\delta \rho}\,,\frac{\delta h}{\delta \rho}\right]\right)
\]
and one can verify that the following  $\operatorname{Diff}(\Bbb{R}^3)-$action is Hamiltonian:
\begin{equation}
\rho(\bx,\bx')\mapsto\frac{\rho(\bbeta^{-1}(\bx),\bbeta^{-1}(\bx'))}{\,\sqrt{\operatorname{det}(\nabla_{\bx}{\bbeta(\bx)}^T\nabla_{\bx'}{\bbeta(\bx')}^T)}\,}
\,. 
\label{DiffeoDensityMatrixAction}
\end{equation}
Given the above left action, the verification detailed in the Appendix \ref{appendix}  shows that its infinitesimal generator may be written as
\begin{equation}\label{DiffeoDensityMatrixInfGen}
u(\rho\,)=-\frac{i}{2\hbar}\left[\{\widehat{u}^k,\widehat{P}_k\}_{+\,},\rho\right]
\end{equation}
where $\widehat{P}_k=-i\hbar\partial_k$ and thus, by using \eqref{Poisson-momap}, one can prove that the corresponding momentum map is given in matrix element notation as
  \begin{align}
\bJ(\rho\,)=\frac12\{\widehat{P},\rho\,\}_+(\bx,\bx)
=:\boldsymbol{\mu}(\bx)
\,.
\end{align}

For the special case of pure states, one verifies that $\rho(\bx,\bx')=\psi(\bx)\psi^*(\bx')$ recovers the momentum map \eqref{Jmomap-Clebsch}. However, in the general case of mixed quantum states, the dynamics of ${\boldsymbol{J}}(\rho\,)$ cannot be expressed only in terms of $\bmu(\bx)$ itself and $D(\bx):=\rho(\bx,\bx)$ as in the case of pure states \cite{Wyatt}. {\color{black}Rather, mixed states lead to a multi-fluid system that is obtained by combining the arguments in Section \ref{1.3} with the relation  \eqref{rhomixture}}. Nevertheless, here we show that the classical limit of mixed state dynamics (as given by the Liouville-von Neumann equation \eqref{LvN}) can be obtained from an exact closure by allowing for the operator $\rho_\text{\tiny$\,$}$ to be sign-indefinite. This should come as no surprise, since classical states violate the uncertainty principle in such a way that the density operator can no longer be positive-definite. The proposed classical closure for the Liouville-von Neumann equation \eqref{LvN} is expressed as
\begin{align}
\rho(\bx,\bx')&=D\Big(\frac{\bx+\bx'}{2}\Big)\exp\!\left[i\frac{m}\hbar 
(\bx-\bx')\cdot\bu\Big(\frac{\bx+\bx'}{2}\Big)\right],
\label{coldfluid}
\end{align}
where $mD(\bx)\bu(\bx)= \boldsymbol{\mu}(\bx)$, as one can show by a direct verification. With the ansatz above, the total energy $\langle\rho,\widehat{H}\rangle$ becomes
$
\langle\rho,\widehat{H}\rangle=
\int ({|\boldsymbol{\mu}|^2}/({2mD}) + 
  DV)\,\de^3 x
  $,
which coincides with the QHD Hamiltonian \eqref{collective2} after dropping the $\hbar^2-$term. Thus, the corresponding equations of motion naturally coincide with the classical hydrodynamic limit \eqref{Restricted-qHcH-eqns}  in terms of Newton's Law, as discussed in Section \ref{Bohmian-paths}. However, it is important to emphasize that, unlike pure states, here the fluid velocity $\bu$ is no longer an exact differential and thus the corresponding hydrodynamic flow preserves the nontrivial circulation $\oint_\gamma \bu(\bx)\cdot\de \bx$, for an arbitrary loop $\gamma$ moving with velocity $\bu$. Then, this produces the vorticity dynamics $\partial_t\bomega=\operatorname{curl}(\bu\times\bomega)$, with   $\bomega=\operatorname{curl}\bu$. In addition, since the probability density is no longer defined in terms of a square-integrable function, the discussion in Section \ref{Bohmian-paths} can be extended by allowing for the multi-particle initial condition \eqref{D-ini}. Then, upon defining $\boldsymbol{q}_a(t)=\bbeta(\boldsymbol{q}_a(0),t)$, the relations \eqref{defns} lead to the Eulerian density $D(\bx,t)=\sum_{a=1}^{\cal N}w_a\delta(\bx-\boldsymbol{q}_a(t))$, where each $\boldsymbol{q}_a(t)$ satisfies Newton's Law for $N$ non-interacting particles. 

We conclude this section by discussing the nature of the closure \eqref{coldfluid} in terms of the Wigner function
$
  W(\bx,\bp)={(2\pi\hbar)^{-3}}\int\rho(\bx+{\by}/2,\bx-{\by}/2) \,e^{-i\frac{\bp\cdot\by}\hbar}\de^3y
$.
Without entering discussions about the Wigner-Moyal formulation of quantum mechanics, we shall simply address the reader to \cite{ZaFaCu2005} and present the closure \eqref{coldfluid} as the operator $\rho$ associated to the following Wigner function:
\begin{align}
  W(\bx,\bp)= D(\bx)\delta({\bp - m\bu(\bx)})
  \,.
  \label{CF-Wigner}
\end{align}
Similar considerations of the cold-fluid closure \eqref{CF-Wigner} have already appeared \cite{JinLi2003} in the context of the semiclassical   limit for pure state dynamics. See also  \cite{BuMoHu2007} and references therein for the use of delta-function closures in  hybrid quantum-classical dynamics.

Finally, we emphasize again that the Wigner function in \eqref{CF-Wigner} does not identify a quantum state. This is analogous to what happens for the quantum harmonic oscillator: in this case, the Wigner-Moyal equation coincides with the classical Liouville equation thereby allowing for delta-function solutions returning classical motion. However, delta-function Wigner distributions do not correspond to quantum states, as their associated density operator is sign-indefinite.

\section{Mean-field model}\label{meanfield-sec}

This section presents the mean-field model, which is based on the factorization \eqref{MF-intro}. Although this model fails to retain correlation effects between nuclei and electrons, it is of paramount importance as the basis of most common models in nonadiabatic dynamics. As we shall see, the geometry of quantum hydrodynamics can be directly applied to this model, thereby leading to the most basic example of hybrid {classical-quantum} dynamics.  

\subsection{The mean-field ansatz}

Here we apply the Euler-Poincar\'e method \cite{HoMaRa1998} to formulate the mean-field model in terms of reduction by symmetry. 
Then, the molecular {wave function} for a nucleus and an electron is written as 
$\Psi:=\Psi(\boldsymbol{r},\boldsymbol{x})\in L^2(\Bbb{R}^3\times\Bbb{R}^3)$.
We think of the coordinate $\boldsymbol{r}$ as that corresponding to the nucleus, while $\boldsymbol{x}$ corresponds to the electron. In physics, a \textit{mean-field ansatz} introduces a {factorization} of the type
\beq
\Psi(\boldsymbol{r},\boldsymbol{x},t)=\chi(\boldsymbol{r},t)\psi(\boldsymbol{x},t)
\,,
\label{MFfact}
\eeq
where $\chi(\boldsymbol{r},t)$ is regarded as the {wave function} describing nuclear dynamics (corresponding to the subsystem we want to treat classically) and both $\psi$ and $\chi$ are {normalized} with respect to the coordinate upon which they depend.  The Hamiltonian operator now takes the form 
$
\widehat{H}=\widehat{H}(\widehat{\Gamma},\widehat{Z})
$,
where we introduce the notation  $\widehat{\Gamma}=(\boldsymbol{r},-i\hbar\nabla_{\!\boldsymbol{r}})$ and $\widehat{Z}=(\boldsymbol{x},-i\hbar\nabla_{\!\boldsymbol{x}})$. Upon recalling the pairing
$
\langle\Psi_1,\Psi_2\rangle=\operatorname{Re}\int\Psi^*_1(\boldsymbol{r},\boldsymbol{x})\Psi_2(\boldsymbol{r},\boldsymbol{x})\,\de^3r\,\de^3x
$,
insertion of the ansatz \eqref{MFfact} into the action principle \eqref{DFVarPrinc}
yields a Lagrangian $L:T\mathscr{H}_n \times T\mathscr{H}_e \rightarrow 
\mathbb{R}$, given by
\begin{align}
L(\Psi,\dot\Psi)&\ =\left\langle\chi\psi,i\hbar\dot{\chi}\psi+i\hbar\chi\dot{\psi}\right\rangle-\left\langle\chi\psi,\widehat{H}\chi\psi\right\rangle
\nonumber
\\
&\
= \langle\psi,i\hbar\dot{\psi}\rangle_\bx
+\langle\chi,i\hbar\dot{\chi}\rangle_\br
-\langle \psi\,,{H}'\psi\rangle_\bx
=:L(\chi,\dot\chi,\psi,\dot\psi)
\label{MFDF1}
\,,
\end{align}
where the second equality uses the natural pairings on the respective Hilbert spaces $\mathscr{H}_n$ and $\mathscr{H}_e$,
and we have introduced the effective Hamiltonian
\beq
\label{EffH1}
{H}'(\widehat{Z})=\int_{\Bbb{R}^3}\chi^*(\boldsymbol{r},t)\widehat{H}\chi(\boldsymbol{r},t)\,\de^3r
\,.
\eeq
{\color{black}As the Lagrangian \eqref{MFDF1} is again of the type Dirac-Frenkel, we  focus  on its corresponding Hamiltonian.} The effective Hamiltonian \eqref{EffH1} can generally be understood as a mapping $\chi\mapsto{H}'(\widehat{Z})$ of the type $\mathscr{H}_n\to L(\mathscr{H}_e)$. This is a mapping from the nuclear Hilbert space, $\mathscr{H}_n$, into 
the space $L(\mathscr{H}_e)$ of linear operators on the electronic Hilbert space. As we shall see, the linear operator ${H}'(\widehat{Z})\in L(\mathscr{H}_e)$ is also Hermitian (self-adjoint).

At this point, a further approximation is often introduced; namely, one assumes that the nuclear dynamics can be treated as classical. This assumption produces a mixed {classical-quantum} system. In what follows, we will introduce a geometric approach which restricts the nuclear evolution to classical particle trajectories.

\subsection{Quantum hydrodynamics and nuclear motion\label{MFsec:nucl}}

In this section, we derive the quantum fluid picture for the mean-field model. We assume
the effective Hamiltonian operator \eqref{EffH1} may be computed from a Hamiltonian operator of the form 
\beq
\widehat{H}=-\hbar^2\Delta_{\br}/2M+V_n(\br)+\,\widehat{\!\cal H}_e+V_I(\br,\boldsymbol{x})
\,,
\label{mario}
\eeq
where $\,\widehat{\!\cal H}_e := -\hbar^2 \Delta_{\boldsymbol{x}}/2m + V_e(\boldsymbol{x})$, while $M$ and $m$ denote the nuclear and electronic masses, respectively. Consequently, upon recalling \eqref{Jmomap-Clebsch}, we have 
\begin{align*}
{H}':=&\ \int_{\Bbb{R}^3}\chi^*(\boldsymbol{r},t)\widehat{H}\chi(\boldsymbol{r},t)\,\de^3r
\\
=&\ \,\widehat{\!\cal H}_e+
\int\!\left[\frac1{2M}\frac{|\bJ(\chi)|^2}{|\chi|^2}+\frac{\hbar^2}{8M}\frac{(\nabla|\chi|^2)^2}{|\chi|^2}+|\chi|^2\big(V_n+V_I(\boldsymbol{x})\big)\right]\de^3r
\,,
\end{align*}
where we have suppressed the $\br-$dependence for convenience of notation. 
Thus, upon denoting $D=|\chi|^2$ and $\bmu=\bJ(\chi)$, the mean-field Hamiltonian functional $h(\bmu,D,\psi):=\langle\psi|{H}'\psi\rangle$ reads 
\begin{align}
\begin{split}
h=
\langle\psi|\,\widehat{\!\cal H}_e\psi\rangle+
\int\!\left[\frac1{2M}\frac{|\bmu|^2}{D}+\frac{\hbar^2}{8M}\frac{|\nabla D|^2}{D}+DV_n+D\langle\psi,V_I(\boldsymbol{x})\psi\rangle\right]\de^3r \,.
\label{MFHamiltonian}
\end{split}
\end{align}
This Hamiltonian functional is a mapping $h:\mathscr{H}_e\times(\mathfrak{X}^*(\Bbb{R}^3)\times \text{Den}(\Bbb{R}^3))\rightarrow \mathbb{R}$, where $\mathfrak{X}^*(\Bbb{R}^3)$ is understood to be the space of {1-form }densities on $\Bbb{R}^3$. At this point, according to the procedure outlined in Section \ref{1.3}, we may perform the partial Legendre transform
$
\bu={\delta h}/{\delta \bmu}=M^{-1}{\bmu}/{D}
$,
{\color{black} and write the mean-field Lagrangian in the following collective, or  reduced, form:}
\begin{multline}
\ell(\psi,\dot{\psi},\bu,D)=\int\!\left[\frac{MD}{2}|\bu|^2-\frac{\hbar^2}{8M}\frac{|\nabla D|^2}{D}-DV_n-D\langle\psi,V_I(\boldsymbol{x})\psi\rangle\right]\de^3r
\\
+\langle\psi,i\hbar\dot\psi-\,\widehat{\!\cal H}_e\psi\rangle
\,.
\label{MFLagrangian}
\end{multline}
The reduced mean-field Lagrangian in \eqref{MFLagrangian} defines a map $\ell: T\mathscr{H}_e \times(\mathfrak{X}(\Bbb{R}^3)\times\text{Den}(\Bbb{R}^3))\rightarrow \mathbb{R}$ which can be regarded as a mixed {hydrodynamic/phase-space} Lagrangian. If the term corresponding to the quantum potential were simply discarded in taking the classical restriction of neglecting $\hbar^2$ in the nuclear dynamics, the reduced mean-field Lagrangian \eqref{MFLagrangian} would become,
\beq\label{joe}
\ell(\psi,\dot{\psi},\bu,D)=\int\!\left[\frac{MD}{2}|\bu|^2-DV_n-D\langle\psi,V_I(\boldsymbol{x})\psi\rangle\right]\de^3r
+\langle\psi,i\hbar\dot\psi-\,\widehat{\!\cal H}_e\psi\rangle
\,.
\eeq
{\color{black}An analogous result could be obtained by following an alternative procedure which would exploit the density operator formalism for the nuclear dynamics, as indicated in Section \ref{densities}. In this case, one would obtain the same Lagrangian \eqref{joe}, although with $\nabla\times\bu\neq0$.}

At this point, one can apply Hamilton's variational principle by taking arbitrary variations $\delta\psi$ and constrained variations \eqref{EPvar} for $\bu$ and $D$. In general, a Lagrangian of this type yields the following Euler-Poincar\'e equations of motion \cite{HoMaRa1998}
{\begin{align}
&(\partial_t +\pounds_{\bu})\frac{\delta \ell}{\delta \bu}=D\nabla\frac{\delta \ell}{\delta D}\label{generalMF1}
\,,\\
&(\partial_t +\pounds_{\bu})D=0\label{generalMF2}
\,,\\
&\frac{\delta \ell}{\delta 
{\psi}}-\partial_t\!\left(\frac{\delta \ell}{\delta 
\dot{\psi}}\right) = 0\label{generalMF3}
\,,
\end{align}
where $\pounds_{\bu}$ denotes the Lie derivative along the vector field $\bu$. {\color{black} These equations take the following forms, upon specializing to the mean-field Lagrangian \eqref{joe}:}
\begin{align}
&M(\partial_t\bu+\bu\cdot\nabla\bu)=-\nabla V_n+\langle\psi|\nabla V_I(\boldsymbol{x})\psi\rangle
\,,\label{MF1}
\\
&\partial_t D+\operatorname{div}(D\bu)=0
\,,\label{MF2}
\\
&i\hbar\dot\psi=\left(\,\widehat{\!\cal H}_e+\int \!DV_I(\boldsymbol{x})\,\de^3r\right)\psi
\label{MF3}
\,.
\end{align}

Again, as explained in  Section \ref{Bohmian-paths},  setting $D(\br,t)=\delta(\br-\bq(t))$ and integrating \eqref{MF1} over space yields  classical trajectories. Eventually, the corresponding classical system reads
\beq
M\ddot{\bq}=-\partial_{\bq} V_n(\bq)- \partial_{\bq}\langle\psi,V_I(\bq,\boldsymbol{x})\psi\rangle
\,,\qquad
i\hbar\dot\psi=\left(\,\widehat{\!\cal H}_e+V_I(\bq,\boldsymbol{x})\right)\psi
\,.
\label{MF-eqnss}
\eeq
This classical restriction preserves the variational structure, whose Lagrangian is now given by
$
  L(\bq,\dot{\bq},\psi,\dot{\psi})= {M}|\dot{\bq}|^2/2 - V_n(\bq) + 
  \braket{\psi, i\hbar\dot{\psi} - (\,\widehat{\!\cal H}_e +V_I(\bq,\boldsymbol{x}))\psi}
$.

Equations \eqref{MF-eqnss} represent the standard mean-field model as it is usually implemented in molecular dynamics simulations \cite{Marx} (although here we have focused on the simplest case of one nucleus and one electron). As we can see in the previous equation, the classical-quantum coupling in this model occurs solely through the interaction potential $V_I$.

Unfortunately, this quantum fluid picture of the mean-field model is not satisfactory in many cases, because the mean-field {factorization} \eqref{MFfact} disregards the classical-quantum correlations between nuclei and electrons. A more advanced model capturing part of these correlation effects will be presented in the next section.

\section{Exact {factorization}}\label{EF-sec}

First appearing in von Neumann's book \cite{vonNeumann} and later developed by Hunter \cite{Hunter}, the following {factorization} ansatz has also been called \textit{exact factorization} in recent work \cite{abedi2010exact,abedi2012correlated,alonso2013comment,abedi2013response}:
\beq
\Psi(\boldsymbol{r},\boldsymbol{x},t)=\chi(\boldsymbol{r},t)\psi(\boldsymbol{x},t;\br)
\,.
\label{ExFact}
\eeq
Here, the electron degree of freedom $\psi$ depends parametrically on the nuclear coordinate $\br$. This means that $\psi$ is a smooth map $\psi\in C^\infty(\Bbb{R}^3,\mathscr{H}_e)$ from physical space to the Hilbert space $\mathscr{H}_e=L^2(\Bbb{R}^3)$ of electronic 
{wave function}s. Furthermore, the {factorization} \eqref{ExFact} invokes the {\it partial normalization condition} (PNC)  
\beq
\int|\psi(\boldsymbol{x},t;\br)|^2\,\de^3x =1\,,
\label{PNC}
\eeq
which as a result of \eqref{ExFact} ensures that 
$
{\int |\Psi(\boldsymbol{r},\boldsymbol{x},t)|^2\de^3x
= |\chi(\br,t)|^2}\,,
$
so that $D(\br,t):=|\chi(\br,t)|^2$ may be interpreted as the nuclear probability density. 


\subsection{Wave functions vs. density operators}
The assumption of exact factorization \eqref{ExFact} in the wave function 
transfers to the density operator, which is then written as 
\beq
\Psi(\boldsymbol{r},\boldsymbol{x})\Psi^*(\boldsymbol{r'},\boldsymbol{x'})
=
\chi(\br)\chi^*(\br')\psi(\boldsymbol{x};\br)\psi^*(\boldsymbol{x}';\br')
\,,
\label{Xrho1}
\eeq
where we have dropped the time-dependence for convenience of notation. 
Then, the physical electron density operator is given by the partial trace, written in
matrix element notation as
\begin{align}
\rho_e(\boldsymbol{x},\boldsymbol{x}')=
&
\iint\!\de^3 r\,\Psi(\boldsymbol{r},\boldsymbol{x})\Psi^*(\boldsymbol{r},\boldsymbol{x'})
=
\int\!\de^3 r\, |\chi(\br)|^2 \,\psi(\boldsymbol{x};\br)\psi^*(\boldsymbol{x}';\br)
\,.
\label{Xrho2}
\end{align}
The corresponding nuclear density operator is
\begin{align}
\rho_n(\br,\br')=
& \iint\!\de^3 x\,\Psi(\boldsymbol{r},\boldsymbol{x})\Psi^*(\boldsymbol{r'},\boldsymbol{x})
=
\chi(\br)\chi^*(\br') \int\!\psi(\boldsymbol{x};\br)\psi^*(\boldsymbol{x};\br')\de^3 x
\,,
\label{Xrho3}
\end{align}
in which we notice that the PNC  \eqref{PNC} does not apply. This means
the quantities $\chi$ and $\psi$ are not true wave functions for the nuclei and electrons (which may not even exist in the presence of {\it decoherence}, i.e. quantum mixing). However, we shall continue to refer to them as such, because they retain certain mnemonic relationships. 
We remark that expectation values of nuclear observables involve integration over
the $\br$-parameters of the electron ``wave functions''. More specifically, the expectation value $\langle A_n\rangle$ for a nuclear observable $A_n(\br,\br')$ is given by (again, in matrix element notation)
$
\langle A_n\rangle := \iint \!{\de^3}r\,{\de^3}r'\!\!\int\! {\de^3}x\,\chi^*(\br')\psi^*(\boldsymbol{x};\br')A_n(\br',\br)\psi(\boldsymbol{x};\br)\chi(\br)
$.
As we shall see, this structure of the nuclear density operator leads to important consequences in the development of the exact factorization theory. 

At this stage, we shall only emphasize that all the relations above also apply naturally  in the context of the Born-Oppenheimer approximation \cite{IzFr2016}, thereby indicating again that the interpretation of nuclear and electronic motion in terms of genuine wave functions needs to be revisited. For example, backreaction effects generated by the presence of $\psi$ in \eqref{Xrho3} can lead to \textit{nuclear decoherence effects} since indeed one has $\rho_n^2\neq\rho_n$. This is a general feature of {classical-quantum} coupling \cite{GBTr18}, which in fact erodes purity in both the classical and the quantum subsystems.

\subsection{General equations of motion}

 In this section, we {generalize} the approach to exact factorization used in \cite{abedi2010exact,abedi2012correlated,alonso2013comment,abedi2013response} by allowing for an arbitrary Dirac Hamiltonian functional $h(\chi,\psi) $ and thereby extending the treatment in Section \ref{1.2-sec}. Thus, inserting the ansatz \eqref{ExFact} in the Dirac-Frenkel Lagrangian 
 and then enforcing the PNC in \eqref{PNC} via a Lagrange multiplier $\lambda({\boldsymbol{r}},t)$ gives
\begin{multline}
L(\chi,\partial_t\chi,\psi,\partial_t\psi, \lambda,\partial_t\lambda)
= \text{Re}\int\! \Big[i\hbar\chi^*\partial_t{\chi}+|\chi|^2\langle\psi|i\hbar\partial_t{\psi}\rangle
\\ + \lambda\big(\|\psi\|^2 - 1\big) \Big]
\,\text{d}^3r
-h(\chi,\psi) 
\,,\label{EFDF1}
\end{multline}
where we have introduced the notation:
$
\langle\psi_1|\psi_2\rangle(\br)
=
\psi_1^\dagger\psi_2(\br)
=
\int \!\psi_1^*(\boldsymbol{x};\br)\psi_2(\boldsymbol{x};\br)\,\de^3x
$,
to denote the natural $L^2$ inner product on $\mathscr{H}_e$. 
%

Naturally, the $\lambda$ equation enforces the PNC, and computing the 
$\chi $ Euler-Lagrange equation yields
  \begin{align}
    &i\hbar\partial_t{\chi} + \braket{\psi|i\hbar\partial_t{\psi}}\chi - \frac{1}{2}\frac{\delta h}{\delta \chi} 
  =0 .\label{EFchiequation}
    \end{align}
    {\color{black}
 Consequently, upon using $\hbar\partial_t|\chi|^2=\operatorname{Im}(\chi^*\delta h/\delta\chi)$, we can compute the $\psi$ equation, as
  \begin{align}
   i\hbar|\chi|^2\partial_t{\psi} + \frac{i}{2}\operatorname{Im}\!\bigg(\chi^*\frac{\delta h}{\delta \chi}\bigg) \psi= &\ 
    \frac{1}{2}\frac{\delta h}{\delta \psi} 
  -\lambda\psi. \label{lambdapsiequation}
 \end{align}
 Upon taking the real part of the inner product of this equation with $\psi$, one obtains 
 that
$
   \lambda =\langle\psi, ({\delta h}/{\delta \psi}) \rangle/2- {|\chi|^2}\braket{\psi,i\hbar\partial_t{\psi}}\,.
$
 Analogously, taking  the imaginary part of the inner product of equation \eqref{lambdapsiequation} with $\psi$ yields the following compatibility condition,
\begin{align}
\text{Im}\left\langle \psi \bigg| \frac{\delta h}{\delta \psi} \right\rangle
= \text{Im}\left(\chi^*\frac{\delta h}{\delta \chi} \right).\label{EFELcondition}
\end{align}
In conclusion, this produces the final form of the $\psi$ equation as
  \begin{align}
  (\mathbbm{1}-\psi\psi^{\dagger})\left(i\hbar\partial_t{\psi} - \frac{1}{2|\chi|^2}\frac{\delta h}{\delta\psi}\right) = 
    0\,,
    \label{EFELpsiprojective}
  \end{align}
 in terms of the notation $\psi^{\dagger} \,\cdot := \braket{\psi| 
  \,\cdot\,}$.
Before closing this section, we remark that the difference between equation \eqref{EFELcondition} and the compatibility condition in \eqref{CompCond1} is its non-zero right-hand side, which arises because the inner product is taken only over the electronic degrees of freedom. That is, equation \eqref{EFELcondition} depends on the nuclear coordinate $\br$.}

\subsection{Local phases and gauge freedom\label{sec:gauge}} 
One may observe that the exact {factorization} \eqref{ExFact} is defined only up to compensating local phase shifts of the nuclear and electronic wave functions. Namely, the replacements
\begin{align}\nonumber
\psi(\bx,t;\br)\to&\  \psi'(\bx,t;\br)=e^{-i\theta(\br,t)}\psi(\bx,t;\br) 
\\
\chi(\br,t)\to&\  \chi'(\br,t)=e^{i\theta(\br,t)}\chi(\br,t) 
\label{EF-gaugefreedom}
\end{align}
leave the EF product wave function $\Psi(\br,\bx,t)=\chi(\br,t)\psi(\bx,t;\br)$ invariant for an arbitrary local phase $\theta(\br,t)$. This is a typical example of {gauge freedom} in a field theory.

To specify the evolution completely, one may either transform to gauge invariant functions, such as the electric and magnetic fields in electromagnetism, or one may fix the gauge by imposing one condition per degree of gauge freedom. 
Gauge fixing is not always the best option, though, because it may obscure physical effects arising due to local breaking of gauge symmetry. An example is the Berry phase, which arises from locally breaking the gauge symmetry of phase shifts in nonrelativistic quantum mechanics \cite{berry1984quantal}.  See also \cite{WilczekShapere1989} for broadly ranging discussions of geometric phases in physics.

The gauge freedom under the compensating local phase shifts in \eqref{EF-gaugefreedom} implies that  
$\langle \psi',i\partial_t\psi'\rangle=\partial_t\theta +\langle \psi,i\partial_t\psi\rangle$.
Hence, one may choose $\theta$ at will (gauge fixing) so as to accommodate any value of $\langle \psi',i\partial_t\psi'\rangle$. For example, one may fix 
$
2|\chi|^{2}\langle\psi|i\hbar\partial_t\psi\rangle=\operatorname{Re}\left\langle\psi|({\delta h}/{\delta\psi})\right\rangle
$, 
so that the $\psi$ equation in \eqref{EFELpsiprojective}  reads 
$
2i\hbar|\chi|^{2}\partial_t\psi={\delta h}/{\delta\psi}-{i}\operatorname{Im}\left\langle\psi|({\delta h}/{\delta\psi})\right\rangle\psi
$.
The same type of gauge was chosen in passing from equation \eqref{ProjSchr1} to equation \eqref {HamEq1}, earlier. 

Another convenient choice  consists in fixing $\langle\psi|i\hbar\partial_t\psi\rangle=0$, so that the $\psi$ equation in \eqref{EFELpsiprojective} becomes 
$
2i\hbar|\chi|^{2}\partial_t\psi={\delta h}/{\delta\psi}-\left\langle\psi|({\delta h}/{\delta\psi})\right\rangle\psi
$.
This gauge is called the \emph{temporal gauge} (or {\it Weyl gauge}) in electromagnetism and it has been adopted recently in \cite{abedi2010exact,agostini2015exact,suzuki2015laser}. We remark that gauge theory is also important in other aspects of chemical physics; for example, see \cite{Littlejohn1} for applications of gauge theory in molecular mechanics.

\subsection{The Hamiltonian functional}
We now return to the Hamiltonian operator $\widehat{H}$ in \eqref{mol-eq}, written as the sum in \eqref{mol-Ham-sum}, 
\begin{align*}
  \widehat{H} &= \underbrace{-\frac{\hbar^2}{2M}\Delta_{\boldsymbol{r}}}_{\hbox{$:= \,\widehat{T}_n$}}
  -\,\underbrace{\frac{\hbar^2}{2m}\Delta_{\boldsymbol{x}}
  + V(\boldsymbol{r},\boldsymbol{x})}_{\hbox{$:= \,\widehat{H}_e$}}\, ,
\end{align*}
where $\widehat{H}_e$ is the electron Hamiltonian operator defined in the Introduction. Henceforth, we will suppress the subscript $\boldsymbol{r}$, and write $\nabla_{\boldsymbol{r}}$ simply as $\nabla$. In this simplified notation, we define the {\it Berry connection} \cite{berry1984quantal} as
\begin{align}
    \boldsymbol{A}(\boldsymbol{r},t)&:= \braket{\psi|-i\hbar\nabla\psi} 
    \, , 
    \label{EFADEF}
\end{align}
where the notation $\bsA(\bsr,t)$ suggests that this quantity plays a role as a gauge field, analogous to the magnetic vector potential in electromagnetism. {Here, we recall that  
$\langle\,\cdot\,|\,\cdot\,\rangle$ denotes the natural $L^2$ inner product on $\mathscr{H}_e$ and  $\|\cdot\|$ denotes the corresponding norm (whose values again depend on the nuclear coordinate 
$\br$).}

We also define the {\em effective electronic potential}, $ \epsilon(\psi,\nabla\psi)$, given by
\begin{align}
 \begin{split} 
   \epsilon(\psi,\nabla\psi)&:=\braket{\psi|\widehat{H}_e\psi} 
    +\frac{\hbar^2}{2M}\|\nabla\psi\|^2 - \frac{\boldsymbol{A}^2}{2M}\\
   &=\braket{\psi|\widehat{H}_e\psi} 
    +\frac{\hbar^2}{2M}
    \left\langle\partial_i\psi,(\mathbbm{1}-\psi\psi^\dagger)\partial_i\psi\right\rangle
    . \label{EFepsilonDEF}
 \end{split} 
\end{align}
The last term in \eqref{EFepsilonDEF} is the trace of the real part of the complex \emph{quantum geometric tensor}  \cite{PrVa1980}
\beq
\label{QGTensor}
Q_{ij}:=\left\langle\partial_i\psi|(\mathbbm{1}-\psi\psi^\dagger)\partial_j\psi\right\rangle=\langle\partial_i\psi|\partial_j\psi\rangle - \hbar^{-2}A_i A_j
\,,
\eeq
where we denote $A_j := \braket{\psi|-i\hbar\partial_j\psi}$. The imaginary part of $Q_{ij}$  is proportional to the Berry curvature {$B_{ij}:= \partial_i A_j - \partial_j A_{i\,}$; namely, $2\hbar{\rm Im}(Q_{ij}) = B_{ij}$} \cite{KoSeMePo2017}.  The emergence of the trace of its real part,
\beq
\label{T=ReQ}
 T_{ij}=\operatorname{Re}(Q_{ij}),
\eeq
 in the electron energy in \eqref{EFepsilonDEF} indicates the geometry underlying the present formulation. 
Notice that the interpretation of $ \epsilon(\psi,\nabla\psi)$ in \eqref{EFepsilonDEF} as an effective electronic potential  departs slightly from that found in the literature, where this quantity is called the \emph{gauge invariant part of the time-dependent potential energy surface} \cite{agostini2015exact,agostini2016quantum,suzuki2015laser}. 

After some manipulation involving completing the square and integration by parts, the Hamiltonian for exact {factorization} in \cite{abedi2010exact,abedi2012correlated} can be expressed as
{\begin{align}\nonumber
  h(\chi,\psi) :=&\  \int\braket{\chi\psi,\widehat{H}\chi\psi}\,\text{d}^3r 
  \\
  =&\ 
  \text{Re}\int  \left(\frac{1}{2M}\,\chi^*(-i\hbar\nabla+ 
  \boldsymbol{A})^2\chi+|\chi|^2\epsilon(\psi,\nabla\psi)\right)
\,\text{d}^3r\, .
\label{EF-Ham1}
\end{align}}
This formula for the EF Hamiltonian agrees with the the chemical physics literature; see, e.g., \cite{suzuki2015laser,agostini2015exact,agostini2016quantum}. 

{\color{black}
For the Hamiltonian \eqref{EF-Ham1}, one computes
\begin{align}
\frac{\delta h}{\delta\chi}=&\ 2(\widehat{T}_n+\epsilon(\psi,\nabla\psi))\chi+\frac{1}{M}\boldsymbol{A}\cdot(\bA-2i\hbar\nabla)\chi
\nonumber
\\
\frac{\delta h}{\delta\psi}=&\ 
2|\chi|^2\widehat{H}_e\psi-2i\hbar^2\frac{\operatorname{Im}(\chi^*\nabla\chi)}{M}\cdot\nabla\psi-i\hbar^2\frac{\operatorname{Im}(\chi^*\Delta\chi)}{M}\psi- \frac{\hbar^2}{M}\operatorname{div}(|\chi|^2\nabla\psi)
\label{FunDer1}
\end{align}
so that the Euler-Lagrange equations \eqref{EFchiequation}-\eqref{EFELpsiprojective} {specialize} to,}
\begin{align}
  &\color{black}i\hbar\partial_t{\chi} = \left(\widehat{T}_n  + \langle \psi | i\hbar\partial_t{\psi} \rangle+\epsilon(\psi,\nabla\psi)\right)\chi +\frac{1}{2M}\boldsymbol{A}\cdot(\bA-2i\hbar\nabla)\chi \label{EFEL1}\,,\\
  &(\mathbbm{1}-\psi\psi^{\dagger})\left[i\hbar\partial_t\psi +\hbar^2\frac{\chi^*\nabla\chi}{M|\chi|^2}\cdot\nabla\psi-\widehat{H}_e\psi+ \frac{\hbar^2}{2M}\Delta\psi  
  \right]=0\,.
  \label{EFEL2}
\end{align}
These equations also agree with the results found in the recent chemical physics literature \cite{abedi2010exact,abedi2012correlated}. In this case, one can show that $\lambda$ vanishes identically (in agreement with \cite{abedi2013response}) and that the compatibility condition \eqref{EFELcondition} is indeed satisfied.

\subsection{Hydrodynamic approach}\label{EFHydrodynamicNuclearSection}
In this section, we again transform into the hydrodynamic picture for the nuclear 
wave function $\chi$. Upon applying the same procedure as in the mean-field case, now denoting  collective fluid variables as $D=|\chi|^2$ 
and $\boldsymbol{\mu}:=\boldsymbol{J}(\chi)=\hbar\text{Im}(\chi^*\nabla\chi)$, 
the  Hamiltonian functional \eqref{EF-Ham1} reads
\begin{align}
h(\boldsymbol{\mu},D,\psi)=
\int\left(\frac{1}{2M}\frac{|\boldsymbol{\mu}+ D\boldsymbol{A}|^2}{D}+ \frac{\hbar^2}{8M}\frac{(\nabla
  D)^2}{D}+ D\,\epsilon(\psi,\nabla\psi)  \right)\text{d}^3r\, \label{EFHydroHamiltonian}.
  \end{align}
At this stage, we will treat the quantity $\psi$ in \eqref{EFHydroHamiltonian} as a parametric variable, whose variations will be taken independently of those for $\bsmu$ and $D$. Applying the partial Legendre transform in this case yields 
$
\bu={\delta h}/{\delta \bmu}=M^{-1}({\bmu+D\bA})/{D}
$,
with $\bA$ defined in terms of $\psi$ in \eqref{EFADEF}, {\color{black}and we write the EF Lagrangian in the following form:}
\begin{multline}
  \ell(\bu,D,\psi, \partial_t{\psi}, \lambda, \partial_t{\lambda}) = \int \bigg[\frac{1}{2}MD |\bu|^2 - D\boldsymbol{A}\cdot \bu- \frac{\hbar^2}{8M}\frac{(\nabla  D)^2}{D} 
  \\ + D\big(\braket{\psi,i\hbar\partial_t{\psi}} - 
  \epsilon(\psi,\nabla\psi)\big)+ \lambda\big(\|\psi\|^2-1\big)\bigg]\text{d}^3r 
  \,.\label{EFHydroL1}
\end{multline}
We now apply Hamilton's variational principle by taking arbitrary 
variations in $\delta\psi$ and $\delta \lambda$, and Euler-Poincar{\'e} variations \eqref{EPvar} for $\bu$ and $D$. 
{As before, a Lagrangian of this type yields general equations of motion 
\eqref{generalMF1}-\eqref{generalMF3},} this time along with a standard Euler-Lagrange equation in $\lambda$, which naturally recovers the PNC. Then, the equation for the electronic {wave function} $\psi$ reads
\beq
(\mathbbm{1}-\psi\psi^{\dagger})\left( i\hbar \partial_t{\psi} + i\hbar \boldsymbol{u}\cdot\nabla\psi - \frac{1}{2D}\frac{\delta F}{\delta \psi}\right) 
  = 0\,,
\label{EFpsiprojective}
\eeq
where we have introduced the functional
\begin{align}
F(D,\psi) := \int \!D \epsilon(\psi,\nabla\psi)\,\text{d}^3r\,,
\label{Fdef}
\end{align}
which we will call the {\it electronic Hamiltonian functional}. Upon making use of the effective electronic potential $ \epsilon(\psi,\nabla\psi)$ defined in \eqref{EFepsilonDEF}, one 
obtains the functional derivative 
\begin{align}
\begin{split}
\frac{\delta F}{\delta \psi}&=D\frac{\partial \epsilon}{\partial \psi} - \text{div}\left(D\frac{\partial \epsilon}{\partial 
\nabla\psi}\right)\\
&=  2D \widehat{H}_e\psi+\frac{i\hbar}{M}D\boldsymbol{A} \cdot \nabla \psi +\frac{i\hbar}{M}\operatorname{div}\!\big(D(i\hbar\nabla+\boldsymbol{A})\psi\big), 
\label{EPphysicalFderiv}
\end{split}
\end{align} 
whose insertion into \eqref{EFpsiprojective} yields
\begin{align}
  (\mathbbm{1}-\psi\psi^{\dagger})\left( i\hbar \partial_t{\psi} + i\hbar  \Big(\boldsymbol{u}-\frac1{2M}\boldsymbol{A}\Big)\cdot\nabla\psi -\widehat{H}_e\psi -\frac{i\hbar}{2MD}\operatorname{div}\!\big(D(i\hbar\nabla+\boldsymbol{A})\psi\big) \right) 
  &= 0 
  \,.
  \label{EFpsiprojectivePhysical}
\end{align}
Again, we obtain a compatibility condition from varying the Lagrange 
multiplier $\lambda$ in \eqref{EFHydroL1}. In this case, the relation \eqref{EFELcondition} becomes
\begin{align}
  \text{Im}\bigg\langle \psi \bigg| \frac{\delta F}{\delta \psi}\bigg\rangle = 0  
  ,
  \label{EFEPcondition}
\end{align}
as one can see by replacing the Hamiltonian  \eqref{EFHydroHamiltonian} in \eqref{EFELcondition}. In addition, one can show that $F$ is invariant under local phase transformations. As we will see in Section \ref{Bohmions}, this local $U(1)$ invariance will ultimately lead to a density matrix formulation of the electronic dynamics.

The Lagrangian \eqref{EFHydroL1} yields the electron dynamics \eqref{EFpsiprojective}, as well as the following Euler-Poincar\'e equations of nuclear hydrodynamics. Specifically, the Euler-Poincar\'e variations  \eqref{EPvar} yield
\begin{align}
&M(  \partial_t + \bu\cdot \nabla)\bu =- \nabla(V_Q + \epsilon) - \boldsymbol{E}   - \bu \times \boldsymbol{B}\label{EFEPELu1} 
\,,\\
 & \partial_t D + \text{div}(D\bu) = 0\,.\label{Dens-eq}
\end{align}
Here, $V_Q$ denotes the nuclear quantum potential \eqref{qPotential-def} for the nuclei, while $\boldsymbol{B}:=\nabla\times\boldsymbol{A}$ is the {\it Berry curvature}, effectively a magnetic field generated by the electrons, and 
$
\boldsymbol{ E}:=-\,\partial_t \boldsymbol{A} - \nabla\langle{\psi|i\hbar\partial_t\psi}\rangle
$
plays the role of an electric field generated by the electrons.  In this analogy, $\hbar$ in the Berry connection defined in equation \eqref{EFADEF} plays the role of the coupling constant (charge) in the electromagnetic force on a charged particle. 

The quantity $\langle{\psi|i\hbar\partial_t\psi}\rangle$ in the definition of $\boldsymbol{ E}$ can be fixed by selecting a particular gauge; {\color{black}for example, possible options are the  {\it temporal gauge}, $\braket{\psi|i\hbar\partial_t\psi}=0$, and the {\it hydrodynamic gauge} $\braket{\psi|i\hbar\partial_t\psi}=\boldsymbol{A}\cdot\bu$}. A more explicit expression of  $\boldsymbol{ E}$ can be found by using \eqref{EFpsiprojective}, thereby leading to the equation
\beq
\boldsymbol{ E}=-\, \partial_t \boldsymbol{A} - \nabla\langle{\psi|i\hbar\partial_t\psi}\rangle
= -\, \bu\times\boldsymbol{B}-\frac1D\left\langle\nabla\psi,\frac{\delta F}{\delta \psi}\right\rangle,
\label{ElField}
\eeq
where we recall the notation $\langle\,\cdot\,,\,\cdot\rangle
= \operatorname{Re}\braket{\,\cdot\, | \,\cdot\,}$ in \eqref{braket} for the real-valued pairing. 
Then, we have
\begin{align}
\begin{split}
\left\langle\nabla\psi,\frac{\delta F}{\delta \psi}\right\rangle-D\nabla\epsilon
=&\ 
\left\langle \nabla\psi, D\frac{\partial \epsilon}{\partial \psi} - \operatorname{div}\left(D\frac{\partial \epsilon}{\partial \nabla \psi}\right)\right\rangle 
-D\nabla\epsilon
\\
=&\   -D\langle\psi,(\nabla\widehat{H}_e)\psi\rangle
  -\partial_j \left\langle D
    \nabla\psi, \frac{\partial \epsilon}{\partial \psi_{,\,j}} 
    \right\rangle
    \\
    =&\ -D\langle\psi,(\nabla\widehat{H}_e)\psi\rangle
-    \frac1M  \partial_j\big(\hbar^2D\langle\nabla\psi,\partial_j\psi\rangle-D\boldsymbol{A} A_j\big).
\end{split}    
    \label{relation}
\end{align}
This may be written in components, in terms of the real part of the quantum geometric tensor $T_{ij}={\rm Re}(Q_{ij})$ in equation \eqref{T=ReQ}, as
$
\left\langle \partial_i\psi\,,({\delta F}/{\delta \psi})\right\rangle-D\partial_i \epsilon
    =    \ -D   \langle\psi,(\partial_i\widehat{H}_e) \psi\rangle
-    M^{-1}  \partial_j \big(DT_{ij}\big)
$.
%
%
Then, the equations \eqref{EFEPELu1}-\eqref{Dens-eq} and \eqref{EFpsiprojective}  take the form,
\begin{align}
&M(  \partial_t  + \bu\cdot \nabla) u_i = 
- \,\partial_i V_Q 
+ \langle\psi,(\partial_i \widehat{H}_e) \psi\rangle 
-  \frac1{MD}  \partial_j\big(DT_{ij}\big)\,,
\label{EFEPELu2} \\
 & \partial_t D + \text{div}(D\bu) =  0\,,
 \label{density}
  \\
 &  (\mathbbm{1}-\psi\psi^{\dagger})\left( i\hbar \partial_t{\psi} 
 + i\hbar  \Big(\boldsymbol{u}-\frac1{M}\boldsymbol{A}\Big)\cdot\nabla\psi -\widehat{H}_e\psi 
 + \frac{\hbar^2}{2MD}\operatorname{div}\!\big(D\nabla\psi\big) \right) 
  = 0\,.
  \label{EFpsi2}
\end{align}
{\color{black}
We emphasize that the equations of motion \eqref{EFEPELu2}-\eqref{EFpsi2} could also be derived from the coupled Schr\"odinger equations \eqref{EFEL1}-\eqref{EFEL2}. Indeed, \eqref{EFEPELu2}-\eqref{density} can be derived from \eqref{EFEL1} by writing $\chi=\sqrt{D}e^{i S/\hbar}$ and by finding the evolution for $\bu=({\nabla S+\bA})/M$. In addition, \eqref{EFpsi2} is equivalent to \eqref{EFEL2}, as it can be verified by replacing \eqref{FunDer1} in \eqref{EFELpsiprojective}.}

%

Notice that equation \eqref{EFEPELu2} does \emph{not} conserve the spatial integral of the nuclear  momentum density,
\beq
MD\bu=\bmu+D\boldsymbol{A}=D(\nabla S+\boldsymbol{A})
\,.
\label{tot-momentum-def}
\eeq
Here, $S$ is the local phase of the wave function $\chi=\sqrt{D}\,e^{iS/\hbar}$ and $D=|\chi|^2$, while $D\boldsymbol{A}$ is part of the nuclear momentum density. This non-conservation of the  hydrodynamic momentum should come as no surprise. In fact, this is already apparent in the  original system \eqref{EFEL1}-\eqref{EFEL2} which, instead, conserves the total  momentum
\beq
\hbar\int\!\!\!\int\!\Psi^*(\br,\bx)(-i\nabla_\br - i\nabla_\bx)\Psi(\br,\bx)\,\de^3 x\,\de^3r=\int\!\left(
\bmu+D\boldsymbol{A}+D\langle\widehat{P}_e\rangle\right)\de^3r
\,,
\label{tot-momentum-redef}
\eeq
{\color{black}where $\langle\widehat{P}_e\rangle = \langle\psi| - i\hbar\nabla_\bx|\psi\rangle$.}
Thus, the total motion has two momentum contributions: one from the $\br$-gradient and the other from the $\bx$-gradient. In particular, the momentum density of the nuclei is given by
$
\hbar\int\!\Psi^*(\br,\bx)(-i\nabla_\br)\Psi(\br,\bx)\,\de^3 x 
= 
\bmu+D\boldsymbol{A}
$
where $\boldsymbol{A}$ is the Berry connection defined in equation \eqref{EFADEF}, and $\boldsymbol{\mu}:=\boldsymbol{J}(\chi)=\hbar\text{Im}(\chi^*\nabla\chi)$.

\subsection{Newtonian limit and Lorentz force} 
The Newtonian limit performed in the chemical physics literature neglects the order quantum potential term in the Lagrangian \eqref{EFHydroL1} and varies the remainder, treating the nuclei as classical particles. The corresponding dynamics may be obtained by neglecting $V_Q$ in \eqref{EFEPELu1} and by replacing $M\bu=\nabla S+\boldsymbol{A}$. One then verifies that this process leads to the following Hamilton-Jacobi equation:
$
{\partial_t S} +{M}^{-1}|\nabla S + \boldsymbol{A}|^2 /2=\langle\psi|i\hbar\partial_t\psi\rangle
- \epsilon(\psi,\nabla\psi) 
$, 
which corresponds to charged particle motion in a Maxwell field, governed by the equation  
\cite{agostini2016quantum}
\beq
M\ddot{\bq}=-\boldsymbol{E}-\dot{\bq}\times\boldsymbol{B}-\nabla\epsilon(\psi,\nabla\psi)\,.
\label{classical}
\eeq
The same result can be obtained by setting $D(\boldsymbol{r},t)=\delta(\boldsymbol{r}-\bq(t))$ in \eqref{Dens-eq} to produce $\dot\bq(t)=\bu(\bq(t),t)$. Then, after multiplying \eqref{EFEPELu1} by $D(\boldsymbol{r},t)$, integration of the delta function over physical space returns the classical equation \eqref{classical} above. 

{\color{black}An important point here is that the customary operation in chemical physics of neglecting the quantum potential term in the Lagrangian \eqref{EFHydroL1} can be problematic. Normally, this step would invoke the limit $\hbar^2\to0$. However, here this process would also lead to discarding the terms ${M^{-1}}({\hbar^2}\|\nabla\psi\|^2 - {\boldsymbol{A}^2})/{2}$ in the effective electronic potential \eqref{EFepsilonDEF}, thereby taking the exact-factorization model into the standard mean-field theory. This crucial issue will be resolved in Section \ref{Bohmions} by performing the exact factorization at the level of the molecular density operator.
}

We should also comment on the Lorentz force appearing in \eqref{classical}. We notice that the combination of this electromagnetic-type force with the potential energy  contribution $\nabla\epsilon$ suggests that the conventional picture of nuclei evolving on potential energy surfaces fixed in space may be oversimplified. As we shall see, despite the claims made in \cite{Suzuki},  the force  $\boldsymbol{E}+\bu\times\boldsymbol{B}$ cannot vanish without requiring major modifications of the electron energy function $\epsilon(\psi,\nabla\psi)$. 
Such modifications would result in singular solution behaviour for the Berry curvature that is unexpected for the exact factorization model. 

In the present context, one may regard the assumption of $\boldsymbol{E}+\bu\times\boldsymbol{B} =0$ as an incompressible magnetohydrodynamic (MHD) approximation of the quantum fluid-plasma equations in \eqref{EFEPELu1} and \eqref{Dens-eq}. 
Setting $D=1$, for which \eqref{Dens-eq} implies ${\rm div}\bu=0$, and taking the curl of \eqref{ElField} yields
\beq
\partial_t \boldsymbol{B} 
= {\rm curl} (\bu\times\boldsymbol{B}) + \left\langle\nabla\psi,\times \nabla \frac{\delta F}{\delta \psi}\right\rangle
\,.\label{3dBerryMHD}
\eeq
Vanishing of the second term on the right side of equation  \eqref{3dBerryMHD} would require a functional relation between $\psi$ and $\delta F/\delta \psi$.
To see the implications of this requirement, we specialize to the 2D case for which the vector $ \boldsymbol{A}(\boldsymbol{r},t):= \braket{\psi|-i\hbar\nabla\psi}$ lies in the plane, so that $\boldsymbol{B}={\rm curl}\boldsymbol{A}= B \boldsymbol{\hat{z}} $ points normal to the plane. Hence, equation \eqref{3dBerryMHD} becomes
\beq
\partial_t B + \bu\cdot \nabla B 
= \boldsymbol{\hat{z}} \cdot  \left\langle\nabla\psi,\times \nabla \frac{\delta F}{\delta \psi}\right\rangle
= J\left(\psi,\frac{\delta F}{\delta \psi}\right)
\label{2dBerryMHD}
\eeq
where $\boldsymbol{\hat{z}} \cdot \nabla f \times \nabla h = J(f,h)$, is the Jacobian between functions $f$ and $h$ on the $(x,y)$ plane. As expected, vanishing of the right hand side of equation \eqref{2dBerryMHD} requires a functional relation between $\psi$ and $\delta F/\delta \psi$. This occurs, for example, in the Ginzburg-Pitaevskii description of Bose-Einstein condensation (BEC) \cite{PiSt2003}, in which case the $B$-equation \eqref{2dBerryMHD} admits singular Berry-curvature solutions  of the form 
$
B (x,y,t) 
= \sum_k \Gamma_k \delta(x-x_k(t)) \delta(y-y_k(t))  
$, {for constants}  $\Gamma_k$ \cite{Fetter}.
The emergence of this type of singular behaviour in the solutions of equation \eqref{2dBerryMHD} is precluded by the dependence on both $\psi$ and $\nabla\psi$ of the electron energy function $\epsilon(\psi,\nabla\psi)$ in equation \eqref{Fdef} for the exact factorization model. 
%

\subsection{Circulation dynamics for the Berry connection\label{Berry-frequency}}
The dynamics of the Berry connection $\bsA(\bsr,t) := \braket{\psi|-i\hbar\nabla\psi}$  in \eqref{EFADEF} governs the circulation of the \textit{nuclear} fluid around closed loops which move with the velocity $\bu = (\bmu/D+\bA)/M$. To see this, we write the motion equation \eqref{EFEPELu2} as the Lie derivative of a circulation 1-form, 
\begin{align}
\begin{split}
M\big( \partial_t + \pounds_u \big) \big( \bu\cdot \de \br \big) 
&= 
-\, \de \left( \frac12 |\bu|^2 + V_Q \right) - \langle\psi,(\de\widehat{H}_e) \psi\rangle
- \,  \frac1{MD}  \partial_j\big(DT_{ij}\big) \de r^i
\\&=
\de \big( \partial_t S + \bu \cdot \nabla S\big)  
+ \big( \partial_t + \pounds_u \big) \big( \bA\cdot \de \br\big)  
\,.\end{split}
\label{transport-mom}
\end{align}
where we have used the relation \eqref{tot-momentum-def} in the second line and $\de$ denotes spatial differential in $\br$. 
Equating the right hand sides of equation \eqref{transport-mom} and integrating around an arbitrary closed loop $\gamma(t)$ moving with the nuclear flow velocity $\bu(\br,t)$ annihilates the exact differential terms and produces the following circulation dynamics for the Berry connection:
    \begin{align}
       \frac{\de}{\de t}\oint_{\gamma(t)} \boldsymbol{A} \cdot \de\boldsymbol{r} 
       =  - \oint_{\gamma(t)} \Big( \langle\psi,(\partial_i\widehat{H}_e) \psi\rangle 
            + \frac1{MD} \partial_j \big(DT_{ij}\big) \Big) \de r^i 
    \,.
    \label{Berry-Frequency}
    \end{align}
This means that the nuclear circulation integral $\oint_{\gamma(t)} \boldsymbol{A} \cdot \de\boldsymbol{r}$, interpreted as the Berry phase obtained by integration around a loop moving with the nuclear fluid, is generated dynamically by an interplay between nuclear and electronic properties. Likewise, the evolution of the Berry curvature $\bB={\rm curl}\bA$ follows by applying the Stokes theorem to relation \eqref{Berry-Frequency}.   
 
Thus, the flux of the Berry curvature through a surface $S$ whose boundary $\partial \Sigma = \gamma(t)$ is a closed loop moving with the nuclear fluid (simply known as the \emph{Berry phase}) satisfies
    \begin{align}
       \frac{\de}{\de t}\int\!\!\!\!\int_{\Sigma} B_{ij}\,{\de}r^j\wedge {\de}r^i
       =  - \oint_{\partial \Sigma}  \Big( \langle\psi,(\partial_i\widehat{H}_e) \psi\rangle 
            + \frac1{MD} \partial_j \big(DT_{ij}\big) \Big) \de r^i 
    \,.
    \label{Berry-CFlux}
    \end{align}
In terms of the real and imaginary parts of the quantum geometric tensor, $Q_{ij}$ in \eqref{QGTensor}, this becomes
    \begin{align}
       2\hbar\frac{\de}{\de t}\int\!\!\!\!\int_{\Sigma} {\rm Im}(Q_{ij}) \,{\de}r^j\wedge {\de}r^i
       =  - \oint_{\partial \Sigma}  \Big( \langle\psi,(\partial_i\widehat{H}_e) \psi\rangle 
       +  \frac1{MD} \partial_j \big(D\,{\rm Re}(Q_{ij})\big) \Big) \de r^i 
    \,.
    \label{QGT-Flux}
    \end{align}
Equation \eqref{QGT-Flux} expresses the quantum geometric mechanics of the correlated nuclear and electronic degrees of freedom in the EF model. Namely, the nuclear probability density $D$ and expectation of the gradient $\nabla \widehat{H}_e$ are coupled dynamically with the real and imaginary parts of the quantum geometric tensor, $Q_{ij}$.


\subsection{Electron dynamics in the nuclear frame}\label{NuclearFrameSection}

So far, we have presented the geometric aspects of the exact factorization model which are currently available in the literature. This section takes a step further to consider the evolution of the electron density matrix. Upon following the arguments in \cite{agostini2016quantum}, we shall write the electron dynamics in the Lagrangian frame moving with the nuclear hydrodynamic flow. 

Recalling the notation $\mathscr{H}_e$ for the electronic Hilbert space, we begin by introducing the group, $C^{\infty}(\mathbb{R}^3, \mathcal{U}(\mathscr{H}_e))$,  of smooth mappings from the physical space into the unitary group of the electronic Hilbert space. Then, we make the following evolution ansatz for the electronic {wave function}: $\psi(t)=(U\psi_0)\circ\eta^{-1}$, or more explicitly
\beq
  \psi(\bx,t;\boldsymbol{r})=
  U(\eta^{-1}(\boldsymbol{r}, t),t)\psi_0(\bx; \eta^{-1}(\boldsymbol{r}, 
  t))
  \,,
\label{unitary-evol}
\eeq
where $\eta$ is the nuclear hydrodynamic path obeying $\dot\eta=\bu(\eta,t)$ and $U(\boldsymbol{r},t)\in C^{\infty}(\mathbb{R}^3, \mathcal{U}(\mathscr{H}_e))$ is a local unitary operator on $\mathscr{H}_e$.
The evolution ansatz \eqref{unitary-evol} results in the following equation for the time evolution of $\psi$:
\begin{align}
 \partial_t\psi +\bu\cdot\nabla\psi&= \xi\psi\,,
\end{align}
where we have defined $\xi := (\dot{U}U^{-1})\circ\eta^{-1}$. Upon substituting these relations into the Lagrangian \eqref{EFHydroL1}, one finds
\begin{align}
  \ell(\bu,\xi,D,\psi) = \int \left[\frac{1}{2}MD |\bu|^2 - \frac{\hbar^2}{8M}\frac{(\nabla 
  D)^2}{D} + D\Big(\braket{\psi,i\hbar\xi{\psi}} - 
  \epsilon(\psi,\nabla\psi)\Big)\right]\text{d}^3r. \label{EFHydroL2}
\end{align}
At this point, we shall prove that the electron energy function $  \epsilon(\psi,\nabla\psi)$ can be written uniquely in terms of the density operator $\rho=\psi\psi^\dagger$ and proceed for the rest of this Section by employing a density matrix description of the electronic dynamics. 
Indeed, we find by direct calculation that
\begin{align}
\begin{split}
\|\nabla\rho\|^2 &=\text{Tr}(\nabla\rho\cdot\nabla\rho)
\\
   &= \text{Tr}\Big(((\nabla \psi)\psi^{\dagger} + \psi (\nabla\psi^{\dagger}) )\cdot((\nabla \psi)\psi^{\dagger} + \psi (\nabla\psi^{\dagger}) 
  )\Big)\\
  &= \braket{\psi|\nabla\psi}^2 + 2\|\nabla\psi\|^2 + \braket{\nabla\psi|\psi}^2 
  \\
  &= 2\operatorname{Tr}T
\,,
\end{split}
\label{minus-calc}
\end{align}
{where the tensor $T$ with components $ T_{ij}=\operatorname{Re}(Q_{ij})$ is the real part of the quantum geometric tensor in equation \eqref{QGTensor}.
Regardless of the minus sign in equation \eqref{QGTensor} the term in the parentheses on the right side of equation \eqref{minus-calc}  is positive, since the left side is positive. Relations resembling} \eqref{minus-calc} also appear in standard quantum mechanics when writing the Fubini-Study metric on the projective Hilbert space $\mathbb{P}_{\!}\mathscr{H}$, see e.g. \cite{Facchi2010}.

Hence, we conclude that the electron energy function \eqref{EFepsilonDEF} may be expressed as,
\beq
\epsilon(\psi,\nabla\psi)=\braket{\rho|\widehat{H}_e} + \frac{\hbar^2}{4M}\|\nabla\rho\|^2
\,,
\label{key-formula}
\eeq
where one defines $\braket{A|B}:=\operatorname{Tr}(A^\dagger B)$ by using the generalized trace. In matrix element notation, one has 
$
\braket{A|B}:=\iint \!A(\bx',\bx)^* B(\bx',\bx)\,\de^3 x\,\de^3x'
$.

The key formula \eqref{key-formula} enables us to write the previous Euler-Poincar\'e Lagrangian equivalently as
\begin{multline}
  \ell(\bu, D, \xi, \rho) =  \int \bigg[\frac{1}{2}MD |\bu|^2-\frac{\hbar^2}{8M}\frac{(\nabla 
  D)^2}{D} 
  \\+ D\bigg(\braket{\rho,i\hbar\xi}  - \braket{\rho|\widehat{H}_e} - \frac{\hbar^2}{4M}\|\nabla\rho\|^2\bigg)\bigg]\text{d}^3r
 \,. \label{EFHydroL3}
\end{multline}
At this point, along with \eqref{EPvar}, one finds the variational relations
\begin{align}
\begin{split}
  \delta \xi &= \partial_t \nu -\bw\cdot\nabla\xi + \bu\cdot\nabla\nu - [\xi,\nu]\,, \\ 
  \delta \rho &= [\nu,\rho]-\bw \cdot\nabla\rho \,,
\end{split}
\label{EFrhovariation}
\end{align}
where we have defined $\nu := (\delta U) U^{-1}\circ\eta^{-1}$. Overall, the relations \eqref{EPvar}, \eqref{EFrhovariation} and the definitions $D=\eta_*D_0$ and $\rho=(U\rho_0U^{-1})\circ\eta^{-1}$ (where for pure states $\rho_0=\psi_0\psi_0^\dagger$) produce the equations of motion for the entire class of reduced Lagrangians of the form $\ell(\bu, D, \xi, \rho)$, with $\bu\in \mathfrak{X}(\Bbb{R}^3)$, {$D\in \text{Den}(\Bbb{R}^3)$}, $\xi\in C^\infty(\Bbb{R}^3,\mathfrak{u}(\mathscr{H}_e))$ and $i\rho\in C^\infty(\Bbb{R}^3,\mathfrak{u}(\mathscr{H}_e))$. Namely, cf. \cite{GBRaTr2012,gay2009geometric,gay2013equivalent,holm2002euler}
\begin{align}
\begin{split}
&  (\partial_t + \pounds_{\bu})\frac{\delta \ell}{\delta \bu} = - \left\langle 
  \nabla\xi, \frac{\delta \ell}{\delta \xi}
  \right\rangle - \left\langle 
  \nabla\rho, \frac{\delta \ell}{\delta \rho}
  \right\rangle + D\nabla\frac{\delta \ell}{\delta D}
  \,,
\\&    (\partial_t + \pounds_{\bu})\frac{\delta \ell}{\delta \xi} - \left[\xi,   \frac{\delta \ell}{\delta \xi}\right] 
    = \left[\frac{\delta \ell}{\delta \rho},\rho\right]\,,
 \\&    (\partial_t + \pounds_{\bu})D = 0\,,
 \\
&     (\partial_t + \pounds_{\bu})\rho = [\xi,\rho]\,.
\end{split}
\label{EFrhoevol}
\end{align}
\begin{remark}[Analogies with complex fluids]
We take this opportunity to make the connection between the hydrodynamic exact {factorization} system and 
previous investigations of the geometry of liquid crystal flows, as found in \cite{GBTr2010,Tronci2012,GBRaTr2012,gay2009geometric,gay2013equivalent,holm2002euler}. 
In this comparison, the electronic {wave function} $\psi(\boldsymbol{r},t)\in C^\infty(\Bbb{R}^3,\mathscr{H}_e)$ is replaced by the director, 
an orientation parameter field $\bn(\boldsymbol{r},t) \in C^\infty(\Bbb{R}^3,S^2)$; the unitary evolution operator $U(\boldsymbol{r},t)\in C^\infty(\Bbb{R}^3,\mathcal{U}(\mathscr{H}_e))$ becomes a rotation matrix $R(\boldsymbol{r},t)\in C^\infty(\Bbb{R}^3,SO(3))$; and one still considers the coupling to the fluid velocity $\bu(\boldsymbol{r},t)$ given by the action of diffeomorphisms $\eta \in \text{Diff}(\mathbb{R}^3)$. 
Indeed, with these replacements, one has a reduced Lagrangian of the same type, $\ell(\bu,\boldsymbol\xi,D,\bn)$, 
and the resulting Euler-Poincar{\'e} equations are equivalent to those in \eqref{EFrhoevol}. 
\end{remark}

For convenience, we rewrite the electronic Hamiltonian functional in \eqref{Fdef} as 
$
  F(D,\rho) = \int D(\braket{\rho|\widehat{H}_e} + M^{-1}{\hbar^2}\|\nabla\rho\|^2/4)\,\text{d}^3r  
$.
Consequently, the fluid velocity equation in \eqref{EFrhoevol} becomes:
\begin{align}
  M(\partial_t + \bu\cdot\nabla)\bu   &= \frac{1}{D}\left\langle \nabla\rho, \frac{\delta F}{\partial \rho} \right\rangle 
  - \nabla(V_Q + \epsilon)
  \,,
\end{align}
which indeed reduces to equation \eqref{EFEPELu2} upon {specializing} to pure states $\rho=\psi\psi^\dagger$. 
Also, we notice the following analogue of relation \eqref{relation}:
\begin{align*}
\left\langle \nabla\rho, \frac{\delta F}{\partial \rho} \right\rangle -D\nabla\epsilon=&\ \left\langle \nabla\rho, D\frac{\partial \epsilon}{\partial \rho} -{\rm div}\left(D\frac{\partial\epsilon}{\partial\nabla\rho}\right)\right\rangle -D\nabla\epsilon
\\
=& -D\langle\rho,\nabla\widehat{H}_e\rangle
-\partial_j\left\langle D \nabla\rho, \frac{\partial \epsilon}{\partial \rho_{,j}} \right\rangle
\\
=&-D\langle\rho,\nabla\widehat{H}_e\rangle
-\frac{\hbar^2}{2M}\partial_j\left\langle D \nabla\rho, \partial_j\rho \right\rangle,
\end{align*}
and the relation for pure states, $\langle\rho,\nabla\rho\rangle=0$, implies $\partial_j\langle D \nabla\rho, \partial_j\rho \rangle=-\partial_j\langle D \rho, \partial_j\nabla\rho \rangle$. 

On the other hand, varying the Lagrangian in \eqref{EFHydroL3} yields the relation {$\delta \ell/\delta \xi = -i\hbar D\rho$}. Upon substituting this relation into the $\xi$ equation in \eqref{EFrhoevol} one finds, via the $D$ and $\rho$ equations, the following simplifying \textit{algebraic relation}
$[i\hbar D \xi - {\delta F}/{\delta \rho},\rho]=0$.
%
The equation in \eqref{EFrhoevol} for density operator $\rho=\psi\psi^\dagger$ now implies the following Liouville-von Neumann  equation,
\begin{align}\label{rhoeqn}
  i\hbar (\partial_t +\bu\cdot\nabla)\rho = \left[\widehat{H}_e-\frac{\hbar^2}{2MD}{\rm div}(D\nabla\rho),\rho\right]
\,.\end{align}

\begin{remark}[Electron decoherence]
Equation \eqref{rhoeqn} will determine the evolution of the electron density matrix defined in \eqref{Xrho2}, 
$
\rho_e(t) := \int D\rho \,\de^3r = \int \tilde\rho \,\de^3r
$.
Namely,
$
i\hbar \dot{\rho}_e(t) 
= 
\int \big[D\widehat{H}_e-{\hbar^2}M^{-1}{\rm div}(D\nabla\rho)/2,\rho\big] \,\de^3r
$.
This result implies that spatially uniform pure initial states (such that $\rho_e^2=\rho_e$) become mixed states as time proceeds. Thus, in agreement with, e.g., \cite{Min}, the exact {factorization} model captures electronic decoherence effects (that is, quantum state mixing) from pure initial states; since the density matrix evolution is no longer unitary.
\end{remark}

Upon collecting equations, now we may specialize the general system \eqref{EFrhoevol}  {to our case} as
\begin{align}
\begin{split}
&  M(\partial_t + \bu\cdot\nabla)\bu   = 
  - \nabla V_Q-\langle\rho,\nabla\widehat{H}_e\rangle -\frac{\hbar^2}{2MD}\partial_j\langle D \nabla\rho,  \partial_j\rho\rangle
  \,,
  \\
  &
    i\hbar (\partial_t +\bu\cdot\nabla)\rho = [\widehat{H}_e,\rho]+\frac{\hbar^2}{2MD}{\rm div}(D[\rho,\nabla\rho])
    \,,\\
    &\partial_t D+{\rm div}(D\bu)=0\,.
\end{split}    \label{final-D-eqn}
\end{align}



\subsection{Hamiltonian structure}
The Hamiltonian structure of equations \eqref{final-D-eqn} can be conveniently rewritten in terms of the quantities
\beq
\bm:=MD\bu\,,\qquad\qquad\tilde\rho:=D\rho\,,
\label{MDrho-var}
\eeq
so that the total energy \eqref{EFHydroHamiltonian} reads
\begin{align}
h(\bm,D,\tilde\rho)=
\int\left(\frac{1}{2M}\frac{|\bm|^2}{D}+ \frac{\hbar^2}{8M}\frac{(\nabla
  D)^2}{D}+ 
\braket{\tilde\rho|\widehat{H}_e} + \frac{\hbar^2D}{4M}\bigg\|\nabla\bigg(\frac{\tilde\rho}{D}\bigg)\bigg\|^2\right)\,\text{d}^3r  
\,. \label{EFHydroHamiltonian2}
  \end{align}
Upon restricting to {pure quantum states}, for which $\langle\rho,\nabla\rho\rangle=0$ and $\langle\tilde\rho,\nabla\tilde\rho\rangle=D\nabla D$, we find
\begin{align}
\frac{\hbar^2D}{4M}\bigg\|\nabla\!\bigg(\frac{\tilde\rho}{D}\bigg)\bigg\|^2=\frac{\hbar^2D}{4M}\bigg\|\frac{\nabla\tilde\rho}{D}-\frac{\nabla D}{D^2}\tilde\rho\bigg\|^2
=
\frac{\hbar^2}{4M}\left(\frac{\|\nabla\tilde\rho\|^2}{D}-\frac{|\nabla D|^2}{D}\right).\label{nablarhocomputation}
\end{align}
Here, the minus sign arises as in the calculation in \eqref{minus-calc}. The term in the parentheses on the right side of this equation is positive, though, since the left side is positive. 

Consequently, for pure quantum states, the Hamiltonian \eqref{EFHydroHamiltonian2} may be expressed as
\begin{align}
h(\bm,D,\tilde\rho)=
\int\left(\frac{1}{2M}\frac{|\bm|^2}{D} -  \frac{\hbar^2}{8M}\frac{|\nabla D|^2}{D}+ 
\braket{\tilde\rho|\widehat{H}_e} + \frac{\hbar^2}{4M}\frac{\|\nabla\tilde\rho\|^2}{D}\right)\,\text{d}^3r  
\,. \label{EFHydroHamiltonian3}
  \end{align}
The appearance of the amplitude of the probability density $D$ in the denominators of the $\hbar^2$ terms in the Hamiltonian in \eqref{EFHydroHamiltonian3} implies that the dynamical effects of the quantum terms in \eqref{EFHydroHamiltonian3} need not decrease, as the squared amplitude of the wave function  decreases. 
  
In terms of the variables $(\bm,D,\tilde\rho)$ defined in \eqref{MDrho-var}, the system of equations in \eqref{final-D-eqn} may be rewritten equivalently, as follows:
\begin{align}
 \begin{split}
 &M\big( \partial_t\bm + {\rm div}(\bu\bm) + (\nabla \bu)^T\cdot \bm \big)  
 = 
       D\nabla V_Q-\langle\tilde\rho,\nabla\widehat{H}_e\rangle -\frac{\hbar^2}{2M}\partial_j\bigg(\frac{\langle  \nabla\tilde\rho,  \partial_j\tilde\rho\rangle}D\bigg)
  \,,
  \\
  &  i\hbar \big(\partial_t \tilde\rho+{\rm div}(\bu\tilde\rho) \big)= [\widehat{H}_e,\tilde\rho]+\frac{\hbar^2}{2M}{\rm div}(D^{-1}[\tilde\rho,\nabla\tilde\rho])\,,
    \\
&    \partial_t D+{\rm div}(D\bu)  =0\,.
 \end{split}
    \label{final-D-eqn2}
\end{align}
A direct verification shows that equations \eqref{final-D-eqn2} may be written in Hamiltonian form ${df}/{dt} = \{f\,,\,h\}$ for a given functional $f(\bm,D,\tilde\rho)$ with the following Lie-Poisson bracket 
\begin{align}
\{f,k\}(\bm,D,\tilde\rho)=&\,\int\!\bm\cdot\left(\frac{\delta k}{\delta \bm}\cdot\nabla\frac{\delta f}{\delta \bm}-\frac{\delta f}{\delta \bm}\cdot\nabla\frac{\delta k}{\delta \bm}\right)\de ^3 r
\\
&\,
-
\int \!D\left(\frac{\delta f}{\delta \bm}\cdot\nabla\frac{\delta k}{\delta D}-\frac{\delta k}{\delta \bm}\cdot\nabla\frac{\delta f}{\delta D}\right)\de ^3 r
\nonumber
\\
&\,
-\int\left\langle i\hbar^{-1}\tilde\rho,\left[\frac{\delta f}{\delta \tilde\rho},\frac{\delta k}{\delta \tilde\rho}\right]+\frac{\delta f}{\delta \bm}\cdot\nabla\frac{\delta k}{\delta \tilde\rho}-\frac{\delta k}{\delta \bm}\cdot\nabla\frac{\delta f}{\delta  \tilde\rho}\right\rangle\de ^3 r
\label{SDP-LPB-EF}
\end{align}
where $h(\bm,D,\tilde\rho)$ is taken to be the Hamiltonian in \eqref{EFHydroHamiltonian3}, and homogeneous boundary conditions are assumed, under integration by parts.

The change of variable $\tilde\rho\to i\hbar\tilde\rho$ shows that this bracket is Lie-Poisson on the dual of the following Lie algebra $\mathfrak{L}$ comprising a direct sum of semidirect product actions:
$\mathfrak{L}=\mathfrak{X}(\Bbb{R}^3)\,\circledS\, \big(C^\infty(\Bbb{R}^3)\oplus\, C^\infty\big(\Bbb{R}^3,\mathfrak{u}(\mathscr{H}_e)\big)\big)
$.
Here, the dual coordinates are $m = \bm\cdot \text{d}\bx \otimes \text{d}^3x \in\mathfrak{X}(\Bbb{R}^3)^* = \Lambda^1(\Bbb{R}^3)\otimes \text{Den}(\Bbb{R}^3)$, $D\in \text{Den}(\Bbb{R}^3)$ and {$i\tilde\rho\in  \mathfrak{u}(\mathscr{H}_e)\otimes\text{Den}(\Bbb{R}^3)$}. Therefore, the Lie-Poisson bracket may be written as 
\beq
\{f,h\}(m,D,\tilde\rho)= -  \left\langle (m,D,\tilde\rho) \,,\, \left[ \frac{\delta f}{\delta (m,D,\tilde\rho)} \,,\,\frac{\delta h}{\delta (m,D,\tilde\rho)}  \right] \right\rangle_{\br},
\label{LPB-brief}
\eeq
where in \eqref{LPB-brief} the angle brackets $\langle\,\cdot\,,\,\cdot\,\rangle_{\br}$ denote $L^2$ pairing in the $r$ coordinates, and the square brackets denote the components of the adjoint action of the semidirect-product Lie algebra $\mathfrak{L}$, whose $\br$-coordinate pairings are given explicitly in equation \eqref{SDP-LPB-EF}.

\section{Density operator factorization and singular solutions}\label{Bohmions}

Our discussion in Section \ref{Berry-frequency} shows that the only contribution to the circulation of the hydrodynamic flow arises from the Berry connection associated to the electronic function $\psi$. This is due to the fact that the hydrodynamic velocity  $M^{-1}\bmu/D=\hbar\operatorname{Im}(\chi^*\nabla\chi)/|\chi|^{2}$ is an exact differential and therefore has zero vorticity.
However, we showed in Section \ref{densities} that this restriction can be relaxed by considering density operators. In the same section we also showed that the classical closure of mixed state dynamics allows for multi-particle trajectories arising from the initial condition \eqref{D-ini}, which in turn is not compatible with the standard QHD definition $D_0=|\chi_0|^2$. In this section we shall  include mixed state dynamics by extending the exact factorization model to a density operator formulation.

\subsection{Factorization of the molecular density operator}
In order to generalize the exact factorization  \eqref{ExFact} to density operators, we recall the relation \eqref{Xrho1} and extend it to consider a  molecular density operator of the form
\begin{align}\label{DMF}
{\rho}_\textrm{\tiny mol}(\br,\br',\bx,\bx')=\rho_n(\br,\br')\psi(\bx;\br)\psi^\dagger(\bx',\br')
\,.
\end{align}
In order to formulate the dynamical equations for such a factorization ansatz, it is convenient to consider a sequence $\{\Psi_k(\br,\bx)\}$ and exploit \eqref{rhomixture} to write
\begin{align}
{\rho}_\textrm{\tiny mol}(\br,\br',\bx,\bx')=\sum_{k=1}^Nw_k\Psi_k(\br,\bx)\Psi_k^*(\br',\bx')
\,,
\label{mixture}
\end{align}
where $\sum_k w_k = 1$ and each $\Psi_k$ satisfies a separate (uncoupled) Schr\"odinger equation with Hamiltonian \eqref{mol-eq}. We remark that  \eqref{mixture} is the equivariant momentum map  for unitary transformations of the $\{\Psi_k\}$ recently studied in \cite{Tronci2018}. Correspondingly, the overall dynamics of $\{\Psi_k\}$ is produced by the variational principle
\beq
L(\{\Psi_k\},\{\dot\Psi_k\})=\sum_{k=1}^Nw_k\!\int\!\!\!\int\!\Big(\hbar\operatorname{Re}\!\big(i\Psi_k^*(\br,\bx)\dot{\Psi}_k(\br,\bx)\big)-\Psi_k^*(\br,\bx)\hat{H}\Psi_k(\br,\bx)\Big)\,\de^3r \de^3 x
\,.
\label{Lag1}
\eeq
At this stage, we consider a $\psi(\bx;\br)$ satisfying  \eqref{PNC}, and we restrict to  the case
$
\Psi_k(\br,\bx)=\chi_k(\br)\psi(\bx;\br)$.
Consequently, the ansatz \eqref{DMF} is recovered by setting $\rho_n(\br,\br')=\sum_kw_k\chi_k(\br)\chi_k^*(\br')$
and the above Lagrangian becomes
\begin{multline}
L(\{\chi_k\},\{\dot\chi_k\},\psi,\dot\psi) =\text{Re}\int\! \bigg[\sum_kw_k\Big(i\hbar\chi_k^*\dot{\chi}_k+|\chi_k|^2\langle\psi|i\hbar\dot{\psi}\rangle\Big) 
\\
+ \lambda\big(\|\psi\|^2 - 1\big) \bigg]
\,\text{d}^3r
-h(\{\chi_k\},\psi) 
\,.
\label{Lag2}
\end{multline}
Here, the Lagrange multiplier enforces $\|\psi\|^2=1$ and the Hamiltonian reads as
$
h(\{\chi_k\},\psi) =\sum_{k}w_k\iint\!\chi_k^*(\br)\psi^*(\bx;\br)\hat{H}(\chi_k\psi)(\br,\bx)\,\de^3 r\,\de^3x
$.
Now, we restrict $\rho_n$ to undergo unitary evolution by writing, for each $k$,
$
\chi_k(\br,t)=U_n(t)\chi_k^{(0)}(\br)
$, 
where $U_n\in\mathcal{U}(\mathscr{H}_n)$ is a time-dependent unitary operator on the nuclear quantum space $\mathscr{H}_n=L^2(\Bbb{R}^3)$.
Then, the Lagrangian above in \eqref{Lag2} can be rewritten in terms of the nuclear density matrix $\rho_n(\br,\br')$ and its diagonal elements $D(\br):=\rho_n(\br,\br)$ as
\beq
\ell(\widehat{\xi}\,,\rho_n,\psi,\dot\psi)=\text{Re}\int\!\Big[ i\hbar\rho_n(\br,\br')\xi(\br',\br)
+D(\br)\langle\psi|i\hbar\dot{\psi}\rangle + \lambda\big(\|\psi\|^2 - 1\big) \Big]\,\de r
-h(\rho_n,\psi) 
\,,
\label{mixednuclearlagrangian}
\eeq
where $\widehat{\xi}:=\dot{U}_nU_n^{-1}$, $\rho_n= U_n\rho_n^{\,\, (0)}U_n^{-1}$, and
the Hamiltonian is:
\begin{align}
  h(\rho_n,\psi) &= \frac{1}{2M}\,\langle(\widehat{P}+ 
  {A})^2|\rho_n\rangle+\int\! D(\br)\epsilon(\psi,\nabla\psi)
\,\text{d}^3r, \label{EF-HamNuclearRho}
\end{align}
in which $\epsilon(\psi,\nabla\psi)$ is the same as in \eqref{EFepsilonDEF} and $\widehat{P}$ denotes the nuclear momentum operator. The Lagrangian \eqref{mixednuclearlagrangian} produces dynamics for $\rho_n$ which is identical to the dynamics for $\chi(r)\chi^*(r')$ emerging from the equation \eqref{EFchiequation}. In this generalized case, the compatibility condition \eqref{EFELcondition} is replaced by
\begin{align}
\text{Im}\left\langle \psi(\br) \bigg| \frac{\delta h}{\delta \psi}(\br) \right\rangle
= 
2\text{Im}\int\!\rho_n(\br,\br')\frac{\delta h}{\delta \rho_n}(\br',\br)\,\de^3 r'.
\end{align}

Now that the variational principle and the Hamiltonian functional are completely characterized, we may proceed by restricting nuclear dynamics to undergo classical motion. To this purpose, we combine the two approaches described in Sections \ref{QHE-Bohmions} and \ref{densities}, by applying the regularization technique after performing the classical closure for density operators.

\subsection{Classical closure and singular solutions}  \label{sec:BohmionsEF}

In following the discussion in Section \ref{densities}, we wish to collectivize the Hamiltonian in terms of the hydrodynamic quantities 
$(\bmu, D)$. This can be achieved by applying the closure \eqref{coldfluid} to $
\rho_n$ in Section \eqref{DMF}, that is 
$
\rho_n(\br,\br')=D(\br/2+\br'/2)\,\exp({i{M} 
(\br-\br')\cdot\bv({\br/2+\br'/2})/\hbar)},
$
with $MD(\br)\bv(\br)=\bmu(\br)$. 
Lengthy but straightforward computations using matrix elements (or, by Wigner transforming, direct applications of Weyl calculus) eventually take the Hamiltonian \eqref{EF-HamNuclearRho} into the hydrodynamic form
\begin{align}
  h(\bmu,D,\psi) &=
\frac{1}{2M}\,
\int \frac{|\bmu+ 
  D{\bA}|^2}{D} \,\de^3 r+\int\! D\epsilon(\psi,\nabla\psi)
\,\text{d}^3r
\,, \label{EFMixedHydroHamiltonian}
\end{align}
which coincides with the Hamiltonian \eqref{EFHydroHamiltonian}, except for the quantum potential term. 
Now however, note that the nuclear hydrodynamic variables $(\bmu,D)$ are no longer given in terms of a unique wavefunction, so that $\operatorname{curl}(M^{-1}\bmu/D)\neq0$.  Thus, equation \eqref{Berry-Frequency} becomes
    \begin{align}
       \frac{\de}{\de t}\oint_{\gamma(t)} \boldsymbol{u} \cdot \de\boldsymbol{r} 
       = - \frac{1}{M} \oint_{\gamma(t)} \Big( \langle\psi,(\partial_i\widehat{H}_e) \psi\rangle 
            + \frac1{MD} \partial_j \big(DT_{ij}\big) \Big) \de r^i 
    \,.
    \label{Berry-Frequency1}
    \end{align}
This means that the dynamics of the nuclear circulation integral $\oint_{\gamma(t)} \boldsymbol{u} \cdot \de\boldsymbol{r}$ is now interpreted as a genuine hydrodynamic Kelvin theorem for the circulation around a loop moving with the nuclear fluid with Eulerian velocity $\boldsymbol{u}:=M^{-1}(\bmu/D+\boldsymbol{A})$. 

At this point, one can Legendre transform the Hamiltonian \eqref{EFMixedHydroHamiltonian} and follow the treatment in Section \ref{NuclearFrameSection} to express the electron function $\psi$ in the nuclear frame. Then, rearranging yields the new Hamiltonian
\begin{align}
    h(\bm,D,\tilde{\rho}) &=
\int \left(\frac{|\bm|^2}{2MD}+ \braket{\tilde\rho|\widehat{H}_e} + \frac{\hbar^2}{4MD}\left({\|\nabla\tilde\rho\|^2}-{|\nabla D|^2}\right)
\right)\,\text{d}^3r 
\,, \label{EFMixedHydroHamiltonian2}
\end{align}
which again coincides with \eqref{EFHydroHamiltonian2} except for the quantum potential, although a similar potential (the last term) with opposite sign is produced  according to the relation \eqref{nablarhocomputation}. The equations of motion associated to the Hamiltonian \eqref{EFMixedHydroHamiltonian2} can now be easily formulated by applying the Lie-Poisson bracket structure \eqref{SDP-LPB-EF}. The last two equations in \eqref{final-D-eqn2} do not change, although the force arising from the quantum potential is modified accordingly in the momentum equation.

Remarkably, Hamiltonian systems with Lie-Poisson brackets of the type    \eqref{SDP-LPB-EF} for the {exact factorization (EF)} model have already been studied and their geodesic motions have been shown to admit singular solutions for certain choices of quadratic Hamiltonians, in \cite{HoTr2009}. Following this previous work, in this Section we will extend the momentum map \eqref{Bohmion-sing-solns1D} which led to the Bohmion singular solutions in Section \ref{QHE-Bohmions}, to a singular solution momentum map for a regularized version of the Lie-Poisson system associated to the Hamiltonian \eqref{EFMixedHydroHamiltonian2}. Again, in analogy with Section \ref{QHE-Bohmions}, here we shall present two different treatments: while the first is uniquely based on the Hamiltonian structure, the second exploits the associated variational principle.

\subsubsection{Hamiltonian regularization in 1D}
For further simplification, we consider a one dimensional nuclear coordinate. Upon considering the Lie-Poisson structure given by the Hamiltonian \eqref{EFMixedHydroHamiltonian2} and the bracket \eqref{SDP-LPB-EF}, we can write the dynamical variables according to the momentum map  given  as follows \cite{HoTr2009}:
\begin{align}
\begin{split}
m(r,t) =   \sum_{a=1}^{\cal N} {p}_a(t) \delta(r-{q}_a(t))
\,,\quad&\quad
D(r,t) =   \sum_{a=1}^{\cal N} w_a(t) \delta(r-{q}_a(t))
\,,\\ &\hspace{-2cm}
\tilde\rho(r,t) =   \sum_{a=1}^{\cal N} \varrho_a(t) \delta(r-{q}_a(t))
\,. 
\end{split}
\label{Bohmion-sing-solns}
  \end{align}
Here, the $D-$equation in \eqref{final-D-eqn2} implies that the $w_a$ with $a=1,2,\dots,N$ in \eqref{Bohmion-sing-solns} are all constant.
In contrast, as we shall see, the $\tilde\rho-$equation for the {regularized} Hamiltonian obtained from introducing smoothed variables in \eqref{EFMixedHydroHamiltonian2} will imply an evolution equation for the coefficients $\varrho_a(t)=\varphi_a(t)\varphi_a^\dagger(t)$ in \eqref{Bohmion-sing-solns}. 
The corresponding singular solutions represent Bohmions in the density matrix formulation.

In addition to the last term in the Hamiltonian \eqref{EFMixedHydroHamiltonian2}, which was already regularized in Section \ref{QHE-Bohmions}, a further barrier to singular solutions is represented by the  term involving the gradient of $\tilde{\rho}$. Hence, we again smoothen our variables by replacing 
$
h(m,D,\tilde\rho)\to h(\bar{m},\Dbar,\bar{\rho})$
 {for} 
$\bar{m}=K*m$,
 $\Dbar=K*D$, 
{and} 
$\bar{\rho}=K*\tilde\rho$,
in the collectivized Hamiltonian \eqref{EFMixedHydroHamiltonian2}. Then, the corresponding {regularized}  Hamiltonian $h_\textit{\tiny REG}$ for the singular solutions is given by
\begin{align}
\begin{split}
h_\textit{\tiny REG}(\{q\},\{p\};\{\varrho\},\{w\}) 
&= \frac{1}{2M} \sum_{a,b} p_ap_b
 \int \frac{K(r-q_a)K(r-q_b)}{\sum_{c} w_cK(r-q_c)} \,\text{d}r 
\\&\quad 
+\frac{\hbar^2}{4M} \sum_{a,b} (\langle\varrho_a |\varrho_b\rangle-w_aw_b)
\int \frac{K'(r-q_a)K'(r-q_b)}{\sum_{c} w_cK(r-q_c)} \,\text{d}r  
\\&\quad 
+ \sum_{a} \langle\varrho_a | \,\widehat{\!\cal H}_e(q_a)\rangle \int K(r-q_a)  \,\text{d}r 
\,.
\end{split}
\label{HamBohm-pqr}
\end{align}
Thus, substitution of the singular solutions \eqref{Bohmion-sing-solns} into the {regularized} Hamiltonian \eqref{EFMixedHydroHamiltonian2} yields the  Hamiltonian \eqref{HamBohm-pqr} in terms of its canonical phase space variables $(q_a,p_a)$ with $a=1,2,\dots,{\cal N}$, augmented as expected by the equation for the matrix $\rho_a$, given by substitution of the last equation in \eqref{Bohmion-sing-solns} into the middle equation of \eqref{final-D-eqn2} as
\beq
i\hbar\dot{\varrho}_a = \left[\frac{\delta h_\textit{\tiny REG}}{\delta \varrho_a},\varrho_a\right]  
=
\left[\bar{H}_e(q_a),{\varrho}_a\right]
+
\frac{\hbar^2}{2M}  \sum_{b} \left[{\varrho_b},{\varrho}_a\right] 
{
\int \frac{K'(r-q_a)K'(r-q_b)}{\sum_{c} w_c K(r-q_c)} \,\text{d}r }
 \,,
\label{R-Bohm-dyn}
\eeq
where we have introduced $\bar{H}_e(q_a)=\int \!H_e(r)\, K(r-q_a)\,\de r$ and we do {\em not} sum on the index $a$.

As for the QHD Bohmions in Section \ref{QHE-Bohmions}, the equivariance of the momentum map \eqref{Bohmion-sing-solns} discovered in \cite{HoMa2005,HoTr2009} implies that the dynamics of $(\{q\},\{p\})$ satisfy the canonically conjugate Hamiltonian equations  
\begin{align}
\dot{q}_a =  \frac{\delta h_\textit{\tiny REG}}{\delta p_a} = u(q_a(t),t)
\quad\hbox{and}\quad
\dot{p}_a =  -\,\frac{\delta h_\textit{\tiny REG}}{\delta q_a} 
\,,\quad\hbox{for}\quad
a = 1,2,\dots,N,
\label{HamcanonBohmion-eqns}
  \end{align}
where $K'$ denotes the derivative of the smoothing kernel $K$ with respect to its argument.  As remarked in Section \ref{QHE-Bohmions} for QHD, these singular solutions for the {regularized} dynamics extend the `peakon' singular solutions for nonlinear wave equations in \cite{CaHo1993,HoONTr2009}. As such, the peakon-like equations in \eqref{Bohmion-sing-solns} do not possess the usual Newtonian form. This is because of the  smoothing which was introduced in the kinetic energy term. 

Again, the term in the canonical equations \eqref{HamcanonBohmion-eqns} arising from summands in $h_\textit{\tiny REG}$ in \eqref{HamBohm-pqr}  proportional to $\hbar^2$ provides extensive, potentially long-range coupling among the singular particle-like solutions, because of the presence of $\Dbar$ in the denominators of these terms in the Hamiltonian $h_\textit{\tiny REG}$ in \eqref{collective2-Bohm}. However, as in the previous section, the limit $\hbar^2\to0$ in the canonical Bohmion equations \eqref{HamcanonBohmion-eqns}  in the density matrix formulation is regular.  



\subsubsection{Lagrangian regularization in 1D\label{sec:LagBohm}}
At this point, we perform the analogous procedure to that in the last part of Section \ref{QHE-Bohmions} now for the above  model obtained by factorizing the molecular density matrix. As before, instead of {regularizing} all terms as in the Hamiltonian approach, we shall {regularize} only the $O(\hbar^2)-$terms.  
We now consider the cold-fluid closure of the Lagrangian \eqref{EFHydroL3}, obtained by Legendre transforming the Hamiltonian \eqref{EFMixedHydroHamiltonian2} to obtain
\begin{align}
  \ell(u, D, \xi, \tilde\rho) &=  \int \bigg(\frac{1}{2}MD u^2+ \braket{\tilde\rho,i\hbar\xi}  - \braket{\tilde\rho|\widehat{H}_e} - \frac{\hbar^2}{4MD}\left({\|{\tilde\rho'}\|^2}-{( 
  D')^2} \right)\bigg)\,\text{d}r\, .
  \label{EFMixedHydroLagrangian}
\end{align}
As a first step, we perform the substitution $D\to\bar{D}$ and $\tilde\rho\to\bar{\rho}$ in the $O(\hbar^2)-$terms, so that the Lagrangian now becomes:
\begin{align}
  \ell(u, D, \xi, \tilde\rho) &=  \int \bigg(\frac{1}{2}MD u^2+ \braket{\tilde\rho,i\hbar\xi}  - \braket{\tilde\rho|\widehat{H}_e} - \frac{\hbar^2}{4M}\left(\frac{\|\bar{\rho}'\|^2}{\bar{D}}-\frac{( 
  \bar{D}')^2}{\bar{D}} \right)\bigg)\,\text{d}r\, .
  \label{EFMixedHydroLagrangian2}
\end{align}
Then in addition to the inital delta function condition on $D$, \eqref{D-ini}, we set the following electronic initial condition:
\beq
\label{rho-ini}
\tilde\rho_0(r_0)=\sum_{a=1}^N\varrho_a^{(0)}\delta(r_0-q_a^{(0)})
\,.
\eeq
Now, the evolution of $\tilde\rho$ in terms of the Lagrangian path $\eta$ and the electronic propagator $U(r_0)$ is given as $
\tilde\rho=\eta_*\hat\rho=
\int\! \hat\rho(r_0,t)\,\delta(r-{\eta}(r_0,t))\,\de r_0
$,
{with} 
$
\hat\rho(r_0,t)=U(r_0,t)\tilde\rho_0(r_0)U^\dagger(r_0,t)
$,
so that \eqref{rho-ini} yields
\[
\tilde\rho(r,t)=\sum_{a=1}^N\varrho_a(t)\delta(r-q_a(t))
\,,\qquad
\text{with}
\qquad
\varrho_a(t):=U(q_a^{(0)},t)\varrho_a^{(0)}U^\dagger(q_a^{(0)},t)
\,.
\]
Here, we set $\varrho_a^{(0)}=\varphi_a^{\scriptscriptstyle (0)}{\varphi_a^{\scriptscriptstyle (0)}}^\dagger$ so that $\varrho_a(t)=\varphi_a(t)\varphi_a^\dagger(t)$ is a projection at all times. 
Furthermore, we evaluate
\[
\int\langle\tilde\rho|i\xi\rangle\,\de r=
\sum_{a=1}^N\langle\varrho_a|i\xi_a\rangle
\,,\qquad
\text{where}
\qquad
\xi_a(t):=\big(\partial_tU(q_a^{(0)},t)\big)U^\dagger(q_a^{(0)},t)
\,,
\]
and we have recalled  $q_a^{(0)}=\eta^{-1}(q(t),t)$ as well as
$
\xi(r)=\big(\partial_tU(\eta^{-1}(r),t)\big)U^\dagger(\eta^{-1}(r),t)
$.
Then, insertion of \eqref{D-ini} and \eqref{rho-ini} in \eqref{EFMixedHydroLagrangian2} yields the following Lagrangian
\begin{multline}
L(\{q\},\{\dot{q}\},\{\varrho\})  =\sum_{a} \Bigg(\frac{Mw_a}2\dot{q}_a^2+\braket{\varrho_a,i\hbar\xi_a}  - \braket{\varrho_a|\widehat{H}_e(q_a)}
  \\
  - \frac{\hbar^2}{4M} \sum_{b} (\langle\varrho_a |\varrho_b\rangle-w_aw_b)  { \int 
  \frac{K'(r-q_a)K'(r-q_b)}
  {\sum_{c} w_c K(r-q_c)} \,\text{d}r }
  \Bigg)
 \,.
\end{multline}
While the Euler-Lagrange equations for the trajectories $q_a$ are obvious, the electronic dynamics is obtained as an Euler-Poincar\'e equation by using the variational relation
$
\delta \xi_a = \partial_t \nu_a - [\xi_a,\nu_a]
$, 
with $\nu_a$ arbitrary and vanishing at the endpoints. As usual, this is obtained by an explicit calculation using the definition of $\xi_a(t)$. Eventually, we obtain the following quantum equation 
\[
i\hbar\dot{\varrho}_a = 
\big[\widehat{H}_e(q_a),{\varrho}_a\big]
+
\frac{\hbar^2}{2M} \sum_{b} \left[{\varrho_b},{\varrho}_a\right]  {   
\int \frac{ K'(r-q_a)K'(r-q_b)}{\sum_{c} w_c K(r-q_c)} \,\text{d}r 
}\,,
\]
which coincides with \eqref{R-Bohm-dyn} except  that $\bar{H}_e$ is now replaced by the unfiltered operator $\widehat{H}_e$. The comparison of solution behaviour between these two regularized Bohmion models in the density matrix formulation will be discussed elsewhere by using computer simulations.

\section{Conclusions}

In this paper, we have  exploited momentum maps to collectivize a sequence of molecular quantum chemistry models for {factorized} nuclear and electronic wave functions, thereby obtaining a sequence of quantum fluid models with shared semidirect-product Lie-Poisson structures.
After reviewing the Born--Oppenheimer product of nuclear and electronic wave functions, we started with mean-field theory, and then passed to a recent development called `exact {factorization}' (EF) for nonadiabatic correlated electron-nuclear dynamics, which has been reported to describe decoherence of pure electron quantum states into mixed states. In the last part, we extended the exact factorization approach to apply for density operators.


In Section \ref{QHE-Bohmions}, we mollified the weakly convergent WKB $\hbar \to 0$ limit by applying a smoothing operator to the quantum variables in the collectivized Hamiltonian for  {regularized} quantum hydrodynamics. This smoothing operation preserved the Hamiltonian structure of the quantum fluid model and it resulted in the discovery of singular delta function solutions called `Bohmions' for smooth quantum fluid Hamiltonians. Depending on which terms are regularized in the Hamiltonian, different sets of Bohmion equations are available.

In the development of the paper, we showed that the Hamiltonian formulation of the collectivized quantum fluid equations for EF possesses the same Lie-Poisson bracket structure as in earlier work on perfect complex fluids (PCF), such as liquid crystals, \cite{GBTr2010,Tronci2012,GBRaTr2012,gay2009geometric,gay2013equivalent,holm2002euler}. The parallel between EF and PCF is that the nuclear fluid velocity vector field Lie-transports both the {nuclear probability density} and the electron density matrix, while the latter also has its own unitary dynamics in the moving frame of the nuclear fluid. 
 This picture was also extended to present a new PCF dynamical model based on the factorization of the molecular density operator.

In the PCF formulation of the nonadiabatic electron problem, smoothing all terms in \eqref{EFMixedHydroHamiltonian2} yields singular momentum maps corresponding to the `peakon' solutions of the well known EPDiff equation \cite{HoMa2005}. In one spatial dimension, this more general class of Bohmions is governed by a countably infinite set of canonical Hamiltonian equations in phase space, in analogy to the solitons for the Camassa-Holm equation \cite{CaHo1993}. The countably infinite phase space system can be truncated to a multi-particle  phase space system at any finite number of Bohmions, because the Hamiltonian dynamics does not create new Bohmions. 
%
%
In fact, the Bohmion collectivized solutions discussed in Sections \ref{QHE-Bohmions} and \ref{sec:BohmionsEF} comprise a semidirect-product version of the class of `peakon' solutions for the CH equation \cite{CaHo1993} which arise from the well-known singular momentum map for the entire class of EPDiff equations \cite{HoMa2005}.

A second approach to Bohmion dynamics was presented in Section \ref{sec:LagBohm}. This approach was developed in the  variational framework by smoothening only the $O(\hbar^2)-$terms in the Lagrangian \eqref{EFMixedHydroLagrangian}. Although the analogy to the peakon solutions of the Camassa-Holm equation  no longer holds entirely, the resulting dynamical system  still consists of a countable set of finite dimensional Hamiltonian equations.

Future work will take further advantage of the analogy between continuum dynamics and the collectivization of quantum dynamics via momentum maps. For example, products of delta functions in different spaces can be introduced, corresponding to Bohmion dynamics for the different {factorized} wave functions of many interacting molecules. This approach is reminiscent of the closure models arising in time dependent Hartree (TDH) theory \cite{Hartree} for quantum dynamics in nuclear physics. Approaches such as these have long been applied in several fields of science, including molecular chemistry, nuclear physics, and condensed matter physics, as well as in celestial mechanics, in hopes of lifting the ``curse of dimensions'' which tends to be ubiquitous in many-body problems \cite{Bonitz}.

\subsection*{Acknowledgements}
{\color{black}The authors are grateful to the referees for their constructive comments that helped improving the presentation of this material. Also, the authors appreciated} stimulating discussions and correspondence with M. Berry, D. Bondar, A. Close, F. Gay-Balmaz, D. Glowacki, S. Jin, T. Ohsawa, L. \'O N\'araigh, D. Ruiz, D. Shalashilin, and S. Wiggins. MF acknowledges the Engineering and Physical Sciences Research Council (EPSRC) studentship grant EP/M508160/1 as well as support from the 2018 Institute of Mathematics and its Applications Small Grant Scheme and University of Surrey FEPS Faculty Research Support Fund. The work by DH was partially supported by EPSRC Standard grant EP/N023781/1 and EPSRC Programme grant EP/P021123/1. The work of CT was  supported by Leverhulme Research Project grant number 2014-112. This material is partially based upon work supported by the National Science Foundation Grant No. DMS-1440140 while MF and CT were in residence at the Mathematical Sciences Research Institute, during the Fall 2018 semester. We all acknowledge the London Mathematical Society Scheme 3 Grant 31633 (Applied Geometric Mechanics Network).

\addtocontents{toc}{\protect\setcounter{tocdepth}{0}}

\appendix
\section{Proof of relation \eqref{DiffeoDensityMatrixInfGen}}\label{appendix}
In this Appendix, we prove formula \eqref{DiffeoDensityMatrixInfGen} for the infinitesimal generator. {\color{black}Instead of considering the left action  in \eqref{DiffeoDensityMatrixAction}, we simplify the treatment here by considering the corresponding right action (given by replacing $\bbeta\to\bbeta^{-1}$). Thus, we begin by considering the following unitary (right) $\text{Diff}(\mathbb{R}^3)$-action:}
\begin{align}\label{gino}
   \Phi_{\bbeta}(\rho)=U_{\bbeta}\,\rho \,U_{\bbeta}^{\dagger}\,,
\end{align}
for the unitary operator (in matrix element notation)
\begin{align*}
  U_{\bbeta}(\bx,\bx')&= \sqrt{\det \nabla_{\bx} \bbeta(\bx)^T}\,\delta(\bx' - 
  \bbeta(\bx))\,.
\end{align*}
At this point we compute the infinitesimal generator from its definition to find
\begin{align}\label{gino-inf}
    \xi(\rho) 
    &= [\hat\xi, \rho]\,.
\end{align}
The matrix elements of $\hat\xi$ can be computed as follows. Upon considering a curve $\boldsymbol{\eta}(t)\in \text{Diff}(\mathbb{R}^3)$ such that $\boldsymbol{\eta}(0)=\boldsymbol{1}$ and $\dot{\boldsymbol{\eta}}(0)=\boldsymbol{u}$, we have
\begin{align*}
   \hat\xi(\bx,\bx')&= \frac{\text{d}}{\text{d}t}\Bigg|_{t=0}(U_{\bbeta}(t))(\bx,\bx')\\
  &=\Bigg[ \frac{\delta(\bx' - 
  \bbeta(\bx,t))}{2\sqrt{\det \nabla_{\bx} \bbeta(\bx,t)^T}} \,\frac{\text{d}}{\text{d}t}
 \big(\!\det \nabla_{\bx} \bbeta(\bx,t)^T\big) + \sqrt{\det \nabla_{\bx} \bbeta(\bx,t)^T}\,\frac{\text{d}}{\text{d}t}\delta(\bx' - 
  \bbeta(\bx,t))\Bigg]_{t=0}\\
    &=\Bigg[\sqrt{\det \nabla_{\bx} \bbeta(\bx,t)^T}\Big(\frac{1}{2} 
  \delta(\bx' - 
  \bbeta(\bx,t))- \dot{\bbeta}(\bx,t)\cdot\nabla_{\bx'}\delta(\bx' - 
  \bbeta(\bx,t)) \Big)\Bigg]_{t=0}\\
    &= \frac{1}{2}\left(\nabla_{\bx}\cdot\bu(\bx)\right)\delta(\bx'-\bx) 
 +\nabla_{\bx}\delta(\bx'-\bx)\cdot \bu(\bx)
 \,.
\end{align*}
The third step uses Jacobi's formula for the derivative of the determinant. 
Next, we show that $\hat\xi={i}{\hbar}^{-1}\{\widehat{u}^k,\widehat{P}_k\}/2$ thereby recovering  \eqref{DiffeoDensityMatrixInfGen}. This is verified as follows:
\begin{align*}
\frac{i}{2\hbar}\{\widehat{u}^k,\widehat{P}_k\}(\bx,\bx') &= \frac{i}{2\hbar}\int 
\widehat{u}^k(\bx,\by)\widehat{P}_k(\by,\bx') + \widehat{P}_k(\bx,\by)\widehat{u}^k(\by,\bx')\,\text{d}^3 
y\\
&= \frac{i}{2\hbar}\int 
i\hbar\delta(\by-\bx')\nabla_{\by_k}\widehat{u}^k(\bx,\by) -i\hbar \delta(\bx - \by) \nabla_{\by_k}\widehat{u}^k(\by,\bx')\,\text{d}^3 
y\\
&=  \frac{i}{2\hbar}\left(i\hbar\nabla_{\bx'_k}\widehat{u}^k(\bx,\bx')-i\hbar  
\nabla_{\bx_k}\widehat{u}^k(\bx,\bx')\right)\\
&=  \frac{i}{2\hbar}\left(i\hbar\nabla_{\bx'_k}\big(u^k(\bx)\delta(\bx-\bx')\big)-i\hbar  
\nabla_{\bx_k}\big(u^k(\bx)\delta(\bx-\bx')\big)\right)\\
&= \frac{1}{2}(\nabla_{\bx}\cdot\bu(\bx))\delta(\bx-\bx') + \bu(\bx)\cdot\nabla_{\bx}\delta(\bx-\bx')
\,,
\end{align*}
where we have used the matrix elements
\[
\widehat{\bu}(\bx,\bx')=\bu(\bx)\delta(\bx-\bx')
\,,\qquad
\widehat{\!\boldsymbol{P}}(\bx,\bx')=-i\hbar\nabla_{\bx}\delta(\bx-\bx')
\,.
\]
{\color{black}Thus we have proved that the infinitesimal generator for the right action \eqref{gino} is indeed given by \eqref{gino-inf}, with $\hat\xi={i}{\hbar}^{-1}\{\widehat{u}^k,\widehat{P}_k\}/2$. Correspondingly, the infinitesimal generator for the left action \eqref{DiffeoDensityMatrixAction} is given by $\xi(\rho) 
    =- [\hat\xi, \rho]$, which then proves \eqref{DiffeoDensityMatrixInfGen}.}


\end{document}